\documentclass[12pt]{article}
\pdfoutput=1 

\usepackage{cancel,slashed}
\usepackage{bbm}
\usepackage{epsf}
\usepackage{amssymb}
\usepackage{amsfonts}
\usepackage{amsmath}
\usepackage{graphicx}
\usepackage{cite}

\usepackage{latexsym}
\usepackage{color}
\usepackage{xypic}
\usepackage{cancel,slashed}
\usepackage[linktocpage]{hyperref}
\usepackage{color}
\usepackage{transparent}

\newcommand{\bmat}{\left(\begin{array}}
\newcommand{\emat}{\end{array}\right)}

\def\p{\partial}
\def\a{\alpha}

\def\b{\beta}
\def\g{\gamma}
\def\d{\delta}
\def\th{\theta}
\def\om{\omega}

\def\-{\hphantom{-}}

\def\s2{\frac{1}{\sqrt2}}

\def\oh{\frac{1}{2}}
\def\beq{\begin{equation}}
\def\eeq{\end{equation}}
\def\beqa{\begin{eqnarray}}
\def\eeqa{\end{eqnarray}}

\def\im{{\rm Im \,}}
\def\re{{\rm Re \,}}

\def\Dsl{\,\raise.15ex\hbox{/}\mkern-13.5mu D} 

\def\CK {{\cal K}}
\def\CM {{\cal M}}

\def\CN {{\cal N}}
\def\CF {{\cal F}}
\def\CL {{\cal L}}

\def\re{\mbox{Re}}
\def\im{\mbox{Im}}

\def\be{\begin{equation}}
\def\ee{\end{equation}}
\def\bea{\begin{eqnarray}}
\def\eea{\end{eqnarray}}
\def\raw{\rightarrow}

\def\IZ{\mathbb{Z}}
\def\IR{\mathbb{R}}

\def\F{{\bf F}}

\def\oh{\frac{1}{2}}
\def\a{{\alpha}}
\def\b{{\beta}}
\def\d{{\delta}}

\def\th{{\theta}}
\def\Lam{{\Lambda}}
\def\lam{{\lambda}}

\def\om{{\omega}}
\def\sig{{\sigma}}
\def\Sig{{\Sigma}}
\def\g{{\gamma}}
\def\G{{\Gamma}}
\def\U{{\Upsilon}}

\def\p{{\partial}}

\def\vec#1{{\overrightarrow{#1}}}

\def\w{{\wedge}}

\def\F{{\mathcal F}}
\def\H{{\mathcal H}}
\def\Q{{\mathcal Q}}

\def\sm2{{\mbox{\small 2}}}

\begin{document}
\pagestyle{plain}

\makeatletter
\@addtoreset{equation}{section}
\makeatother
\renewcommand{\theequation}{\thesection.\arabic{equation}}
\pagestyle{empty}
\rightline{IFT-UAM/CSIC-14-051}
\vspace{0.5cm}
\begin{center}
\LARGE{{U(1) mixing and D-brane linear equivalence}
\\[10mm]}
\large{Fernando Marchesano,$^1$ Diego Regalado$^{1,2}$ and Gianluca Zoccarato$^{1,2}$ \\[10mm]}
\small{
${}^1$ Instituto de F\'{\i}sica Te\'orica UAM-CSIC, Cantoblanco, 28049 Madrid, Spain \\[2mm] 
${}^2$ Departamento de F\'{\i}sica Te\'orica, 
Universidad Aut\'onoma de Madrid, 
28049 Madrid, Spain
\\[8mm]} 
\small{\bf Abstract} \\[5mm]
\end{center}
\begin{center}
\begin{minipage}[h]{15.0cm} 

Linear equivalence is a criterion that compares submanifolds in the same homology class. We show that, in the context of type II compactifications with D-branes, this concept translates to the kinetic mixing between U(1) gauge symmetries arising in the open and closed string sectors. We argue that in generic D-brane models such mixing is experimentally detectable through the existence of milli-charged particles. We compute these gauge kinetic functions by classifying the 4d monopoles of a compactification and analyzing the Witten effect on them, finding agreement with previous results and extending them to more general setups. In particular, we compute the gauge kinetic functions mixing bulk and magnetized D-brane U(1)'s and derive a generalization of linear equivalence for these objects. Finally, we apply our findings to F-theory SU(5) models with hypercharge flux breaking.

\end{minipage}
\end{center}
\newpage
\setcounter{page}{1}
\pagestyle{plain}
\renewcommand{\thefootnote}{\arabic{footnote}}
\setcounter{footnote}{0}


\tableofcontents


\section{Introduction}
\label{s:intro}

Despite the progress made in the construction of realistic and semi-realistic 4d string theory models \cite{thebook}, one of the main challenges that remains for string phenomenology is to obtain model-independent, testable predictions. Lacking a principle that selects one set of string vacua over the rest, a good chance to achieve a predictive framework is to address those scenarios that are the most generic in the plethora of semi-realistic string vacua, and to compute the most robust testable quantities associated to them. 

In general, the most robust quantities of a string compactification are those of topological nature, which at low energies translate into basic field theory objects like the gauge group and chiral matter content of the 4d effective action. In practice, building a string model with a semi realistic gauge sector involves engineering a compactification whose topology yields the chiral content of the Standard Model of Particle Physics (SM), plus possibly a hidden gauge sector necessary to fulfill the consistency conditions of the theory. Given this basic setup, one then proceeds to compute the many parameters that determine the physics of the SM and beyond, with the expectation that the string model will provide a natural explanation for their values as measured in current and future experiments. 

In this sense, a good opportunity to achieve a predictive framework is to focus on those parameters that are more robust from the field theory viewpoint, and which turn out to depend on very few compactification data beyond the topological choices mentioned above. Typical examples of this are holomorphic quantities in 4d supersymmetric theories, like superpotentials or gauge kinetic functions. For instance, one may naturally achieve hierarchical rank 3 holomorphic Yukawas by considering F-theory GUT vacua satisfying a few topological relations \cite{cchv09,fimr12,fmrz13}. Another promising direction is to consider the gauge kinetic function mixing the hypercharge with an extra $U(1)$ symmetry \cite{Hattori:1993zu,Dienes:1996zr,Lukas:1999nh,Lust:2003ky,Abel:2003ue,jl04,Blumenhagen:2005ga,Abel:2006qt,agjkr08,Grimm:2008dq,Goodsell:2009xc,Gmeiner:2009fb,Goodsell:2010ie,Bullimore:2010aj,Cicoli:2011yh,Williams:2011qb,gl11,kt11,cim11,Honecker:2011sm,Goodsell:2011wn,Honecker:2012qr,Shiu:2013wxa}, since these extra $U(1)$'s arise generically in string constructions, and there are a number of ongoing experiments that could detect the related signatures. 

\begin{figure}[htb]
  \centering
  \def\svgwidth{0.6\columnwidth}
  \resizebox{0.75\textwidth}{!}{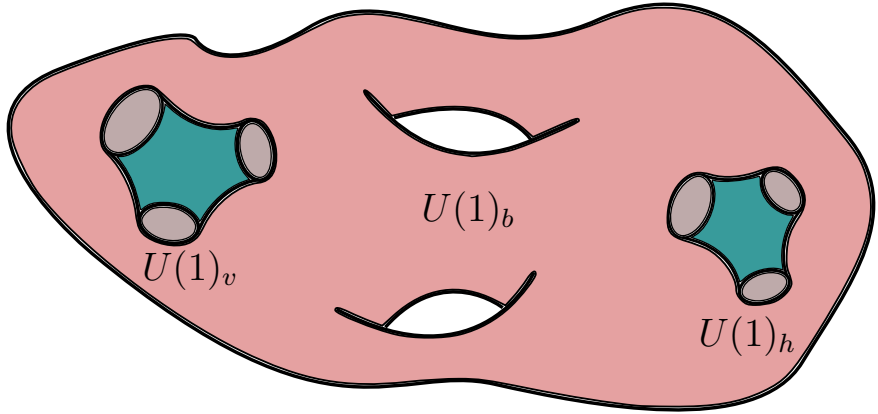}
  \caption{D-brane realization of the milli-charged scenario.}
  \label{fig}
\end{figure}

Particularly interesting is the so-called milli-charged scenario, in which light particles charged under a hidden $U(1)_h$ obtains a small electric charge due to the kinetic mixing of $U(1)_h$ with the hypercharge \cite{Holdom:1985ag}. Following the recent discussion in \cite{Shiu:2013wxa}, one can achieve such effect by considering a string compactification with two different gauge sectors. The first sector contains the SM and in particular the hypercharge $U(1)_Y$, while the other sector contains a hidden gauge group with a second massless $U(1)_h$. One can easily visualize this setup in terms of a standard class D-brane models (c.f. figure \ref{fig}), where the visible sector is localized in a set of D-branes wrapping $p$-cycles in a region of the compactification manifold $\CM_6$, and the hidden sector arises from a second D-brane set located in a different region. The fact that the two sets of D-branes are internally separated from each other guarantees that there are no light particles charged under the visible and gauge sectors simultaneously. Moreover, the gauge kinetic mixing $\chi_{vh}$ between the visible and hidden $U(1)$ will vanish at tree level. Nevertheless, there will be massive particles charged under $U(1)_v =U(1)_Y$ and $U(1)_h$ which, when integrated out, will generate one-loop corrections to the mixing (see e.g. \cite{agjkr08}). Finally, upon diagonalization of the gauge kinetic terms, light matter charged under $U(1)_h$ will acquire a small hypercharge proportional to $\chi_{vh}$ \cite{Holdom:1985ag}.

While this is a rather general scenario, obtaining precise results depends crucially on the computation of threshold corrections to the gauge kinetic mixing, which is technically very difficult beyond simple toroidal orbifold compactifications. Moreover, generically the quantity $\chi_{vh}$ will decrease for large separations of the visible and hidden sectors and small values of the string coupling, and so in this regime its value could be quite small. 

In the following we will describe a slight variation of this scenario resulting into an alternative mechanism to generate milli-charged particles. As we will see below, these new contributions can be computed by a simple geometric formula describing tree-level local quantities, and so their effect could be more important than the one just described.

\subsection{RR photon mixing and linear equivalence}

The variation comes from assuming the presence of a further $U(1)$ gauge symmetry that does not arise from the D-brane degrees of freedom, but rather from the Kaluza-Klein reduction of the Ramond-Ramond closed string sector of the theory. Such bulk $U(1)$'s, dubbed RR photons in \cite{cim11}, exist for generic choices of the compactification manifold $\CM_6$ and are natural sources of hidden $U(1)$ gauge symmetries, because the only particles charged under them are extremely heavy: namely D-branes wrapped on internal cycles and point-like in 4d. Despite their hidden nature, these RR photons will mix kinematically with the D-brane $U(1)$'s \cite{jl04,gl11,kt11,cim11}. However, since there is no light particle charged under them there is naively no measurable effect arising from their mixing with the hypercharge.\footnote{Still, in models with low energy supersymmetry hidden $U(1)$ gauginos may mix with the MSSM neutralinos and this could lead to new signatures at the LHC  \cite{Feldman:2006wd,Feldman:2009wv,Ibarra:2008kn,Arvanitaki:2009hb}. } Let us nevertheless consider the existence of a RR photon $U(1)_b$ living in the bulk and its mixing with the D-brane photons $U(1)_v$ and $U(1)_h$ of figure \ref{fig}. After normalizing the gauge bosons we recover an effective Lagrangian of the form
\be
\CL_{4d}\,\supset \, -\frac{1}{4} \sum_{i=v,h,b} F_{\mu\nu}^{(i)}F^{(i)\, \mu\nu} + \oh \left(\chi_{vb} F_{\mu\nu}^{(v)}F^{(b)\, \mu\nu} +\chi_{hb} F_{\mu\nu}^{(h)}F^{(b)\, \mu\nu} \right)
\label{lagcan}
\ee
where we have assumed that $\chi_{vh}$ is negligible, although it can be easily reincorporated. The two mixing terms can be eliminated by performing the change of basis
\be
\left(
\begin{array}{c}
{A}^{\prime (v)} \\ {A}^{\prime (h)} \\ {A}^{\prime (b)}
\end{array}
\right)
\,=\, 
\left(
\begin{array}{ccc}
\sqrt{\frac{1- \chi^2_{vb} -\chi^2_{hb}}{1- \chi^2_{hb}}} & 0 &  0 \\
-\frac{\chi_{vb} \chi_{hb}}{\sqrt{1- \chi^2_{hb}}} & \sqrt{1- \chi^2_{hb}} & 0 \\
-\chi_{vb} & -\chi_{hb} & 1
\end{array}
\right)
\left(
\begin{array}{c}
{A}^{(v)} \\ {A}^{(h)} \\ {A}^{(b)}
\end{array}
\right)
\label{shift}
\ee
and so, even if we started with no mixing $\chi_{vh}$, the shift in the D-brane hidden gauge boson $A^{(h)}$ induces a hypercharge on the light matter charged under $U(1)_h$ with a suppression factor $\d_{vh}^{\rm eff}= \frac{\chi_{vb} \chi_{hb}}{\sqrt{1- \chi^2_{vb}- \chi^2_{hb}}}$ compared to the SM particles. 
This effect is to be compared with the factor $\d_{vh} = \frac{\chi_{vh}}{\sqrt{1- \chi^2_{vh}}}$ that would arise if we only had a non-vanishing mixing between the two D-brane $U(1)$'s, and naively it seems to be more suppressed than the latter. Nevertheless, as we will see the kinetic mixing between closed and open string $U(1)$'s is not a one-loop suppressed effect, and also that $\d_{vh}^{\rm eff}$ is independent of the relative separation between visible and hidden D-brane sectors. Hence it could be a comparable or even stronger effect than the one considered in the standard milli-charge scenario. 

Motivated by this observation, the aim of this present paper is to perform a systematic study of the kinetic mixing between open and closed string $U(1)$'s in type II compactifications, extending the previous analysis in \cite{cim11}. Rather than (\ref{lagcan}) we will consider the following $U(1)$ action
\be\label{ac4d}
S_{4d,\,U(1)}=-\frac{1}{2\kappa_4^2}\int_{\mathbb R^{1,3}}  \re (f_{pq}) F_2^p\,\w *_{4}F_2^q+\im (f_{pq}) F_2^p\,\w\, F_2^q 
\ee
where $\kappa_4^2=\frac{l_s^2}{4\pi}$, and $l_s=2\pi\sqrt{\a'}$ is the string length.\footnote{In our conventions all the $p$-form potentials are dimensionless (except for the D$p$-brane gauge field) and the field strengths have dimension 1. By $p$-form dimension we mean the  dimension of its components.} 
This formulation encodes the kinetic mixing and the $\theta$ angles of the theory in terms of the gauge kinetic function $f_{pq}$, a protected quantity in 4d supersymmetric theories. While in the following we will not consider any particular scale of SUSY breaking, we will assume that supersymmetry is at least restored at the compactification scale. This will not only guarantee the stability of our constructions, but also that $f_{pq}$ is a holomorphic function of the 4d chiral fields arising below that scale. It will moreover imply that, fixed the few topological data that describe the scenario of figure \ref{fig}, the kinetic function mixing open and closed string $U(1)$'s will depend in relatively few compactification data. 

In fact, we will find that the kinetic mixing $\re f_{pq} = 4\pi \frac{\chi_{pq}}{g_pg_q}$ is closely related to the mathematical concept of linear equivalence. In brief, for an open string $U(1)$ to be massless we need to satisfy certain topological conditions regarding the $p$-cycles wrapped by the D-branes, roughly speaking that a certain linear combination of $p$-cycles is homologically trivial \cite{cim11}. Given such combination of $p$-cycles it is possible to draw a $(p+1)$-chain $\Sigma$ connecting them. The requirement of linear equivalence can then be formulated by asking that the integrals of certain harmonic bulk $(p+1)$-forms over $\Sigma$ vanish \cite{Hitchin99}. This is not a topological condition, in the sense that it depends on the embedding of the $p$-cycles inside their homology class, but it is a rather simple and robust quantity as it involves integrals of harmonic forms over slices of the internal manifold $\CM_6$. As we will see, such integrals over the chain $\Sigma$ are nothing but the gauge kinetic function mixing open and closed string $U(1)$'s, in agreement with our expectations that these protected quantities should be easier to compute than many other 4d effective couplings, which is a promising starting point to draw model-independent predictions out of them. 

The paper is organized as follows. In section \ref{IIAmix} we analyze the mixing between open and closed string $U(1)$'s in type IIA models of intersecting D6-branes. We derive the expression for the mixing by means of the Witten effect applied to the open string monopoles of the compactification, which we classify in terms of relative homology groups. We describe the relation of this kinetic mixing with the linear equivalence of 3-cycles, and provide an expression to compute it in terms of $p$-form integrals over the compactification manifold. In section \ref{IIBmix} we extend this picture to type IIB compactifications with magnetized D7-branes. Because these are equivalent to bound states of D-branes of different dimension, the concept of linear equivalence is more complicated and we need to use the machinery of generalized geometry to properly formulate the kinetic mixing. We finally apply these results to F-theory $SU(5)$ models where the hypercharge mixes with closed string $U(1)$'s. We draw our conclusions in section \ref{conclu}.

Several technical details have been relegated to the appendices. Appendix \ref{Leq} reviews the definition of linear equivalence and several results relevant for our main discussion. Appendix \ref{ap:Taub} analyzes in detail the M-theory lift a set of parallel D6-branes. Appendix \ref{gh} contains the basic definitions of generalized homology relevant for the computations on the main text and Appendix \ref{cal} spells out the details of one of them. 

\section{$U(1)$ kinetic mixing for intersecting D6-branes}
\label{IIAmix}

In this section we discuss the kinetic mixing between $U(1)$'s in type IIA orientifold compactifications. In particular, we consider models made up of D6-branes wrapping special Lagrangian cycles (sLags) of Calabi-Yau three-folds, and describe the kinetic mixing of open string $U(1)$'s with RR $U(1)$'s. We derive the expression for such mixing by means of the Witten effect, recovering the results of \cite{cim11} from a purely type IIA perspective. We then point out the relation between open string $U(1)$'s and the relative cohomology of the compactification manifold, which allows to write down a simple supergravity formula for the kinetic mixing between open and closed string $U(1)$'s. Finally, we discuss the relation between open-closed kinetic mixing and linear equivalence of cycles (see appendix A). 

\subsection{Type IIA Calabi-Yau orientifolds with D6-branes}

Let us consider an orientifold of type IIA string theory on $\mathbb R^{1,3}\times \mathcal M_6$ with $\mathcal M_6$ a Calabi-Yau 3-fold. The orientifold projection is obtained by modding out by the action $\Omega_p(-1)^{F_L}\sigma$ where $\Omega_p$ is the worldsheet parity, $F_L$ is the space-time fermion number for the left-movers and $\sigma$ is an antiholomorphic involution of $\mathcal M_6$ acting as $z_i\,\rightarrow\,\bar z_i$ on local coordinates, which introduces O6-planes. Therefore, the action of the involution on the K\"ahler form $J$ and holomorphic 3-form $\Omega$ of $\mathcal M_6$ is given by
\be
\sigma J=-J,\qquad \sigma \Omega =\bar \Omega.
\ee
The supersymmetry conditions for a D6-brane wrapping a 3-cycle $\pi$ in $\mathcal M_6$ are
\be\label{sLag}
J|_{\pi}=0,\qquad \rm{Im\,} \Omega|_{\pi}=0.
\ee
Since the D6-brane charge lies in the homology group $H_3(\mathcal M_6,\mathbb Z)$, 
cancellation of the total charge in the compact internal space can be written as
\be
\sum_\a N_\a([\pi_\a]+[\pi_\a^*])-4[\pi_O]=0
\label{RRtad}
\ee
where $\a$ is an index that runs over the set of branes, $N_\a$ is the total number of branes on $\pi_\a$ and $\pi_\a^*=\sigma\pi_\a$ is the cycle wrapped by the orientifold image of $\a$. Finally, $\pi_O$ is the fixed locus of the involution $\sigma$ where the O6-planes lie and the factor $-4$ is due to the O-plane RR charge, assumed to be negative. 

The 4d massless spectrum that arises from the closed string sector of the compactification can be computed upon dimensional reduction of the 10d type IIA supergravity action, and is given in terms of harmonic forms on $\mathcal M_6$. It is then useful to introduce a basis of harmonic forms of definite parity under the involution $\sigma$,
\be\nonumber
\begin{array}{ccc}
&\sigma\rm{-even}&\sigma\rm{-odd}\\
\rm {2-forms}\quad&\omega_i\quad i=1,\dots,h_+^{1,1}\quad&\omega_{\hat i}\quad \hat i=1,\dots,h_-^{1,1}\\
\rm {3-forms}\quad&\a_I\quad I=0,\dots, h^{1,2} \quad&\b^I\quad I=0,\dots, h^{1,2}\\
\rm {4-forms}\quad&\tilde\omega^{\hat i}\quad \hat i=1,\dots,h_-^{1,1}\quad&\tilde\omega^i\quad i=1,\dots,h_+^{1,1}\end{array}
\ee
normalized such that
\be
\int_{\mathcal{M}_6}\omega_i\,\w\,\tilde \omega^j=l_s^6\d_i^j,\qquad\int_{\mathcal{M}_6}\omega_{\hat i}\,\w\,\tilde \omega^{\hat j}=l_s^6\d_{\hat i}^{\hat j},\qquad\int_{\mathcal{M}_6}\a_I\,\w\,\b^J=l_s^6\d_I^J.
\label{intersection}
\ee
For instance, in order to reduce to 4d the RR forms $C_3$ and $C_5$ one can expand them as
\bea
\label{c3O}
C_3&=&A^i_1\,\w\,\omega_i+\re(N^I)\a_I \\
\label{c5O}
C_5&=&C_{2,I}\,\w\,\b^I+ V_{1,i} \wedge \tilde\omega^i
\eea
where we have taken into account the intrinsic parity of $C_3$ and $C_5$ (respectively even and odd) under the orientifold action. One then obtains that $C_3$ gives rise to $h_{1,1}^-$ axions $\re(N^I)$ and to $h_{1,1}^+$ gauge bosons $A_1^i$, while $C_5$ contains their 4d dual degrees of freedom. 

A convenient basis of harmonic $p$-forms is given by those that have integer cohomology class, that is whose integrals over any $p$-cycle are integer numbers. In particular we will choose the 2-forms $\omega_i$, $\omega_{\hat{i}}$ such that $[\omega_i] \in H^2_+(\CM_6, \IZ)$, $[\omega_{\hat{i}}] \in H^2_-(\CM_6, \IZ)$.\footnote{In general $H^2(\CM_6, \IZ)$ may not decompose as $H^2_+(\CM_6, \IZ) \oplus H^2_-(\CM_6, \IZ)$, but the latter may only be a sublattice of the former. In this case one should introduce appropriate factors of $2$ in (\ref{intersection}). To simplify in the following we will assume that (\ref{intersection}) holds even if $[\omega_i]$, $[\tilde \omega^i]$, $[\omega_{\hat{i}}]$, $[\tilde \omega^{\hat{i}}]$ all belong to integer cohomology.} This automatically implies that a D2-brane wrapping a 2-cycle $\Lam_2$ will have integer electric charges under the RR $U(1)$'s. More precisely, a D2-brane wrapping a 2-cycle $\Lam_2^j$ whose class is Poincar\'e dual to $[\tilde \omega^j] \in H^4_-(\CM_6, \IZ)$ will have electric charge $\delta^{ji}$ under the RR $U(1)$ generated by $A_1^i$. Another consequence of this choice is that the gauge kinetic mixing for RR $U(1)$'s takes the simple form \cite{gl04}
\be
f_{ij} \, =\, -\frac{i}{2l_s^4}  \int_{\CM_6}  J_c  \wedge \omega_i \wedge \omega_j \, =\, -\frac{i}{2} \CK_{ij\hat{k}} T^{\hat{k}}
\label{clclmix}
\ee
where we have defined the complexified K\"ahler moduli $T^{\hat{k}}$ by
\be
J_c  \, \equiv \, B_2 + i J \, =\, l_s^2\, T^{\hat{k}} \omega_{\hat{k}}
\label{Jc}
\ee
while  
\be
\CK_{ij\hat{k}} \,  \equiv \, \frac{1}{l_s^6}\int_{\CM_6} \omega_i \wedge \omega_j \wedge \omega_{\hat{k}} \quad \quad 
\ee
are the triple intersection numbers, which in this basis are simply integers. 

Regarding the open string sector of the compactification, the 4d massless spectrum that arises from a single D6-brane $\a$ wrapping a 3-cycle $\pi_\a$ is given by
\bea
\label{WLD6}
A_1^{\a}&=&\frac{\pi}{l_s}(A_1^{4d,\a}+\th_\a^j\,\zeta_j)\\
\phi_\a&=&\phi_\a^jX_j
\label{posD6}
\eea
where $A_1^{4d,\a}$ is a 4d gauge vector field (we will henceforth suppress the superscript 4d)\footnote{The $\frac{1}{l_s}$ is introduced to keep $A_1^{4d,\a}$ and $\th_\a^j$ dimensionless and the factor of $\pi$ for later convenience. The field $\phi_a^j$ is related to the normal coordinate by $y_a^j=\frac{l_s}{2}\phi_a^j$ so it is also dimensionless.}. Moreover, $\th_\a^j$ are the components of the corresponding Wilson line moduli with
\be
\frac{\zeta_j}{2\pi}\in\rm{Harm}^1(\pi_\a,\mathbb Z)
\label{zetadef}
\ee
and $\phi_\a^j$ are the D6-brane position moduli, namely the components of a normal deformation of the brane preserving the sLag conditions (\ref{sLag}) with
\be
X_i\in N(\pi_\a)\ {\rm such \ that} \ \mathcal L_{X_i}J=\mathcal L_{X_i}\rm{Im }\Omega=0
\ee
where $\mathcal L_{X_i}$ is the Lie derivative along $X_i$. These two scalar fields together form a 4d complex modulus, namely
\be
\Phi_\a^j=\th_a^j+\lam_i^j\phi_\a^i
\ee
with $\lam_i^j$ a complex matrix relating $\{\zeta_j\}$ and $\{X_i\}$ and defined by
\be
\iota_{X_i}J_c|_{\pi_\a}=\lam_i^j\zeta_j
\label{lambda}
\ee
where $J_c$ is the complexified K\"ahler form (\ref{Jc}). It is straightforward to generalize this spectrum to the case of a stack of $N_\a$ D6-branes wrapping $\pi_\a$, so that the 4d gauge group is given by $U(N_\a)$ and $\Phi_\a^j$ transform in its adjoint representation. Finally, 4d chiral multiplets may arise from the transverse intersections of $\pi_\a$ with its orientifold image $\pi_\a^*$ as well as with other 3-cycles wrapped by the remaining D6-branes of the compactification \cite{thebook,reviews}.

\subsection{Separating two D6-branes}

In order to discuss kinetic mixing between open an closed string $U(1)$'s let us follow \cite{cim11} and first consider type IIA strings on $\mathbb R^{1,3}\times\mathcal M_6$, without any orientifold projection, and suppose that we have two D6-branes $a$ and $b$ wrapping the same sLag 3-cycle $\pi_a=\pi_b$. This leads to a gauge group $U(2)$ in 4d, which breaks down to $U(1)_a\times U(1)_b$ when these two 3-cycles are separated. However, only a linear combination of these two $U(1)$'s remains massless at low energies, while the other one becomes massive due to the St\"uckelberg mechanism.

Indeed, let us consider the CS action for a single D6-brane wrapping a 3-cycle $\pi_\a$, which is obtained from a 3-cycle $\pi$ after a small normal deformation of the form $\phi_\a = \phi_\a^j X_j$.\footnote{Such normal deformation should be small enough so that $\pi$ and $\pi_\a$ have the same topology, and in particular the same number of non-trivial 1-cycles.} We have that (see \cite{gl11,kt11,cim11} for further details)
\bea
\label{csa}
S^\a_{CS}& \supset & \mu_6 \int_{\mathbb R^{1,3}\times \pi_\a} \left [\mathcal F^\a_2\,\w\,C_5+\frac{1}{2}\mathcal F^\a_2\,\w\,\mathcal F^\a_2\,\w\,C_3\right ]\\ \nonumber
& = &  \mu_6 \int_{\mathbb R^{1,3} \times \pi} e^{\CL_{\phi_\a}}
  \left [\mathcal F^\a_2\,\w\,C_5+\frac{1}{2}\mathcal F^\a_2\,\w\,\mathcal F^\a_2\,\w\,C_3\right ]\
\eea
with $\mu_6=\frac{2\pi}{l_s^7}$ the D6-brane charge, $\CL_{\phi_\a} = \phi_\a^j \CL_{X_j}$ the Lie derivative along such deformation and $\mathcal F^\a_2=\frac{l_s^2}{2\pi}F^\a_2+B_2$. In the absence of orientifold projection the RR 5-form potential $C_5$ has the expansion
\be
\label{c5}
C_5\, =\, C_{2,I}\,\w\,\b^I+\tilde C_2^I\,\w\,\a_I+V_{1,i} \wedge \tilde\omega^i
\ee
where now $i = 1, \dots, h^{1,1}$ runs over all harmonic 2-forms in $\CM_6$. We now consider two 3-cycles $\pi_a$ and $\pi_b$ that are deformations of $\pi$ and wrap a D6-brane on each of them. The full CS action then contains the following piece
\be
\label{csn}
S^a_{CS}+S^b_{CS}\, \supset\, \frac{\pi}{l_s^6}\left [\int_{\mathbb R^{1,3}}(F_2^a+F_2^b)\,\w\,C_{2,I}\int_{\pi}\b^I+ \int_{\mathbb R^{1,3}}(F_2^a+F_2^b)\,\w\,\tilde C_2^I\int_{\pi}\a_I\right ]
\ee
where we have used that the integrals of $\b^I$, $\a_I$ only depend on the homology class of the 3-cycle, and in particular that $\int_{\pi} e^{\CL_{\phi_\a}} \beta^I = \int_{\pi} \beta^I$, same for $\a_I$. For a non-trivial $[\pi]$ some of these integrals will be non-vanishing, and so the combination $U(1)_a+U(1)_b$ will develop a $BF$ coupling and therefore a St\"uckelberg mass, while the orthogonal combination
\be
\label{abU1}
U(1)_{(a-b)}\, =\, \oh \left[U(1)_a-U(1)_b\right]
\ee
will remain massless. 

We can now read off the kinetic mixing of (\ref{abU1}) with the RR $U(1)$'s from the remaining terms of $S^a_{CS}+S^b_{CS}$. For this it is useful to consider the following expansion for the RR potentials 
\bea\label{c3}
C_3&=&A^i_1\,\w\,\omega_i+\dots\\
\label{c5d}
C_5&=& \tilde A^i_1\,\w\,*_6 \omega_i+\dots 
\eea
where the dots represent terms that do not contain 4d gauge bosons. This new expansion for $C_5$ is chosen so that $*_4F^{{\rm RR},i}_2 \equiv *_4dA_1^i  = d\tilde A_1^i$. Plugging the expansion (\ref{c5d}) into (\ref{csa}) and projecting into the combination $F_2^{(a-b)}\equiv \oh [F_2^a-F_2^b]$ we obtain
\be
\label{csn2}
S^a_{CS}+S^b_{CS}\, \supset\, - \frac{\pi}{2l_s^5} \int_{\mathbb R^{1,3}} F_2^{(a-b)} \wedge * F^{{\rm RR},i}_2 \int_{\pi} \left( \iota_{\phi_a}J -  \iota_{\phi_b}J\right) \wedge\, \omega_i + \dots
\ee
where we have only kept terms linear in the deformations $\phi_\a$. Here we have used that $\CL_\phi = d \iota_\phi + \iota_\phi d$ and that $*_6 \omega_i = b_i J^2 -J \wedge \omega_i$ with $b_i = (3 \int_{\CM_6} \omega_i \wedge J^2)/(2 \int_{\CM_6} J^3)$. Using the definition (\ref{lambda}) we can recast this result as
\be
\re f_{i (a-b)}\, =\, \frac{1}{4l_s^3}(\phi^k_a - \phi^k_b)\, \im (\lam^j_k) \int_{\pi} \zeta_j \wedge \omega_i
\label{ref2}
\ee
where the basis of 1-forms $\{\zeta_j\}$ is defined as in (\ref{zetadef}). 

The imaginary part of $f_{i (a-b)}$ is obtained from plugging (\ref{c3}) into (\ref{csa}), which gives 
\be
\label{csn3}
S^a_{CS}+S^b_{CS}\, \supset\, \frac{\pi }{2l_s^5} \int_{\mathbb R^{1,3}}F_2^{(a-b)} \w\, F^{{\rm RR},i}_2  \int_{\pi} \left[(\th_a^j-\th_b^j) \zeta_j + \left( \iota_{\phi_a}B -  \iota_{\phi_b}B\right) \right]\wedge \omega_i + \dots
\ee
where we have again integrated by parts and kept terms linear in the deformations. Comparing to the general expression (\ref{ac4d}) we conclude that
\be
\im f_{i (a-b)}\, =\, - \frac{1}{4l_s^3} \left[ (\th_a^j-\th_b^j) + (\phi^k_a - \phi^k_b)\, \re (\lam^j_k) \right] \int_{\pi} \zeta_j \wedge \omega_i
\label{imf2}
\ee
Adding this result to (\ref{ref2}) we obtain
\be\label{kinm}
f_{i(a-b)}=- \frac{i}{4l_s^3} (\Phi_a^j-\Phi_b^j) \int_{\pi}\zeta_j \wedge \omega_i
\ee
which as expected is a holomorphic function of the D6-brane moduli. Notice that the mixing vanishes for $\Phi_a^j=\Phi_b^j$, which corresponds to the case where the two branes are on top of each other and the gauge group enhances to $SU(2)$. 

In order to arrive at (\ref{kinm}) we assumed that $\pi_a$ and $\pi_b$ are obtained from deforming the same 3-cycle $\pi$. As a result, they are not only in the same homology class but are also homotopic. However, since the vanishing of the St\"uckelberg mass depends only on the homology of the cycles and not on their homotopy class, one would like to have an expression for the kinetic mixing that applies in the general case. In \cite{cim11}  such a formula was found to be 
\be\label{kinb}
f_{i(a-b)}=- \frac{i}{2l_s^4}\int_{\Sigma}\left (J_c+ \frac{l_s^2}{2\pi} \tilde F_2^{(a-b)}\right )\,\w\,\omega_i
\ee
where $\Sigma$ is a 4-chain such that $\partial \Sigma=\pi_a-\pi_b$, and $\tilde F_2^{(a-b)}$ is such that
\be\label{wldef}
\int_{\Sigma} \tilde F_2^{(a-b)}\,\w\,\omega_i\equiv \left [\int_{\pi_a}A_1^a\,\w\,\omega_i-\int_{\pi_b}A_1^b\,\w\,\omega_i\right ].
\ee
It can be easily shown that this 4-chain expression reproduces (\ref{kinm}) for homotopic branes. Moreover, following \cite{cim11} one can see that the kinetic mixing needs to be of the form (\ref{kinb}) by performing the M-theory lift of these compactifications. In the next section we will arrive at (\ref{kinb}) from yet a different viewpoint, without using any M-theory lift.

Notice that the open-closed kinetic mixing vanishes if
\be
\int_{\pi}\omega_i\,\w\,\zeta_j = \int_{\rho_j} \omega_i = 0
\ee
where the 2-cycle $\rho_j \subset \pi$ is Poincar\'e dual to $\zeta_j$. This is true only if none of the 2-cycles of $\pi$ is non-trivial in $\CM_6$. By the results of \cite{Hitchin99}, this is equivalent to saying that the 3-cycles $\pi_a$ and $\pi_b$ are linearly equivalent. Hence, in the present case linear equivalence of D-branes translates into a vanishing kinetic mixing with RR photons, as advanced in the introduction. In the following sections we will see how this statement can be generalized to more involved D-brane configurations. 

\subsubsection*{Orientifolding}

Let us now include the effect of the orientifold projection. Because in this case $C_5$ has the expansion (\ref{c5O}), instead of (\ref{csn}) we obtain
\bea
\label{cs}
S_{CS}^a+S_{CS}^{a^*}+S_{CS}^b+S_{CS}^{b^*} & \supset &\frac{1}{2}\frac{\pi}{l_s^6}\int_{\mathbb R^{1,3}}(F_2^a+F_2^b)\,\w\,C_{2,I} \left[\int_{\pi}\b^I- \int_{\pi^*}\b^I\right]\\ \nonumber
& = & \frac{\pi}{l_s^6} \int_{\mathbb R^{1,3}}(F_2^a+F_2^b)\,\w\,C_{2,I} \int_{\pi}\b^I
\eea
where the extra factor of $1/2$ arises due to the orientifold projection, and we have used the fact that $F_{\a^*}=-F_\a$ for the 7d gauge field. As a result, now $U(1)_a + U(1)_b$ will develop a St\"uckelberg mass if and only if $[\pi] \neq [\pi^*]$. 

Let us assume that this is the case and compute the kinetic mixing for the massless $U(1)$ (\ref{abU1}), for which one can obtain expressions similar to the unorientifolded case. Indeed, we have that
\be
S_{CS}^a+S_{CS}^{a^*}+S_{CS}^b+S_{CS}^{b^*} \, \supset \,\frac{\pi}{2l_s^5}
\int_{\mathbb R^{1,3}} F_2^{(a-b)}\,\w\,F_2^{{\rm RR},i}\cdot (\th_a^j-\th_b^j) \int_{\pi} \zeta_j \wedge \omega_i
\ee
from where we can deduce that the kinetic mixing again takes the form (\ref{kinm}). In terms of a 4-chain formula we would again arrive to (\ref{kinb}), with the only difference that now $\Sigma$ is defined by $\p \Sigma = \pi_a - \pi_b - \pi_a^* + \pi_b^*$.\footnote{In this case the 4-chain sigma can be divided into two pieces as $\Sigma =\Sigma_{ab} - \Sigma_{ab}^*$, where $\p \Sigma_{ab} = \pi_a - \pi_b$ and $\Sigma_{ab}^*$ is its orientifold image.}

On the other hand, for $[\pi]=[\pi^*]$ both $U(1)_a$ and $U(1)_b$ remain massless. One should then be able to write the kinetic mixing of each $U(1)_\a$ with the RR $U(1)$'s individually. Indeed, one finds that the expression analogous to (\ref{kinb}) is 
\be
f_{i\a}=-\frac{i}{2l_s^4} \int_{\Sigma_\a}\left (J_c+\frac{l_s^2}{2\pi} \tilde F_2^\a\right )\,\w\,\omega_i
\label{kin22}
\ee
where $\Sigma_\a$ is defined in the covering space $\CM_6$ and satisfies $\p\Sigma_\a'=\pi_\a- \pi^*_\a$.  

\subsection{Kinetic mixing via the Witten effect}

Let us now describe an alternative derivation for the kinetic mixing formula (\ref{kinb}), based on the Witten effect \cite{Witten79}. This effect is the fact that for $U(1)$ gauge theories that break $CP$ and contain magnetic monopoles, the latter acquire an electric charge proportional to the $CP$ breaking term. For theories whose $CP$ violating effect is a $\th$-term this electric charge can be computed exactly, namely
\be
Q^E=-\frac{\th}{2\pi}e.
\ee
This can be generalized to theories that contain multiple $U(1)$'s and whose action is described by (\ref{ac4d}). The lattice of charges is then \cite{bjk09}
\bea
\label{chargel}
Q^E_I&=&n^e_I-\im f_{IJ}\, n_J^m\\\nonumber
Q^M_I&=&n_I^m
\eea
where $I=1,\dots, K$ runs over the set of massless $U(1)$'s and {$n_e^I\in\mathbb Z/2$}, $n_m^I\in \mathbb Z$ are the charges that appear in the action when we include this particle. In other words, including a particle with charges $(n^e_I,n^m_I)$ amounts to consider the action 
\be
S=S_{4d,\,U(1)}+\frac{4\pi}{l_s}Q^E_I\int_W A^I+\frac{4\pi }{l_s} \tilde Q^M_J \int_W \tilde A^J
\label{actcharge}
\ee
where $W$ is the worldline of the 4d particle, $d\tilde A^I = *_4 dA^I$  and $\tilde Q^M_J = - \re f_{JI} Q^M_I$.

The basic strategy to determine the open-closed $U(1)$ mixing will be to consider the 4d magnetic monopoles of D6-brane $U(1)$'s and compute their electric charges $Q_i^E$ under a closed string $U(1)_i$. Given the above facts, such electric charge should be proportional to the imaginary part of the gauge kinetic function computed in previous sections. Moreover, by looking to the couplings of these monopoles to other closed string $U(1)$ magnetic generators $\tilde A^i_1$ we will also be able to obtain the real part of the mixing. As before we will first consider the case of parallel D6-branes without O6-planes and subsequently include the orientifold projection.

Considering again the system of two homotopic D6-branes wrapping $\pi_a$ and $\pi_b$, the 4d monopole with unit charge under $U(1)_{a-b} = \oh \left[U(1)_a- U(1)_b \right]$ is given by a D4-brane wrapping $W\times\Sigma$, where $W$ is the worldline of the monopole in 4d and $\partial \Sigma=\pi_a-\pi_b$. In order to compute the electric charge under the closed string $U(1)$'s we can dimensionally reduce the D4-brane CS action to obtain
\be
S^{D4}_{CS}\supset \mu_4 \int_{W\times \Sigma}C_3\,\w\,\mathcal F^{D4}=\mu_4 \int_{W\times \Sigma} A_1^i\,\w\,\om_i\,\w\, \F^{D4}=\frac{4\pi }{l_s} Q^E_i \int_{W}A_1^i
\ee
where the electric charges are given by
\be\label{ch}
Q^E_i=\frac{1}{2l_s^4} \int_{\Sigma}\F^{D4}\,\w\,\om_i.
\ee
This term is precisely (minus) the imaginary part of the mixing in (\ref{kinb}). In particular, the field strength in $\F^{D4}=B_2+\frac{l_s^2}{2\pi}F_2^{D4}$ is such that $F_2^{D4}|_{\pi_\a}=F_2^{\a}$, as the monopole must interpolate between the two D6-brane configurations. More precisely, $F_2^{D4}$ is the curvature of a line bundle on $\Sigma$ such that on its boundary $\p \Sig$ it reduces to the line bundles on the corresponding D6-brane. Such line bundle is nothing but the Wilson lines $A_1^\a$, and so one recovers (\ref{wldef}) by simply identifying $\tilde F_2^{(a-b)}$ with $F_2^{D4}$. Notice that (similarly to the 4-chain $\Sigma$) there are many $F_2^{D4}$ that have the appropriate boundary conditions. As we will see in the next section the line bundle extension $F_2^{D4}$ appears naturally in the context of generalized complex geometry.

Let us now consider the coupling of this monopole to the 4d dual vector boson $\tilde A_1^i$ that appears in the expansion (\ref{c5d}) of the RR potential $C_5$. Looking at the appropriate term on the D4-brane CS action we obtain
\be
S^{D4}_{CS}\supset \mu_4 \int_{W\times \Sigma}C_5=- \mu_4 \int_{W\times \Sigma} \tilde A_1^i\,\w\,J\,\w\,\omega_i= - \frac{4\pi}{l_s} \re f_{i(a-b)}  \int_{W}\tilde A_1^i
\ee
where we have again used that $*_6 \omega_i = b_i J^2 -J \wedge \omega_i$ and assumed that $b_i =0$, which will be automatically satisfied in the orientifold case. We then have that the real part of the mixing is given by
\be
\re f_{i(a-b)} =\frac{1}{2l_s^4}\int_{\Sigma}J\,\w\,\omega_i
\ee
which as expected reproduces (\ref{kinb}). Alternatively, we could have interchanged the role of the dual vector bosons $A_1^i \leftrightarrow \tilde A_1^i$ and applied the Witten effect  to obtain this result. 

Notice that in this derivation we do not need to assume that the two 3-cycles $\pi_a$ and $\pi_b$ are homotopic, nor that they relatively close to a reference 3-cycle $\pi$. The only requirement is that they are homologous so the St\"uckelberg mass vanishes and a 4-chain $\Sigma$ exists. This then provides an alternative way to derive the expression (\ref{kinb}) from first principles without having to perform the lift to M-theory. Finally, it is straightforward to extend this derivation to the orientifold case and again obtain (\ref{kinb}), except that now the 4-chain $\Sigma$ and the field strength $F_2^{D4}$ should connect the 3-cycles $\pi_\a$ and $\pi_\a^*$. 

\subsection{General case}

It is clear how to generalize these results to arbitrary D6-brane configurations. Indeed, let us consider $K$ stacks of D6-branes, each stack containing $N_\a$ D6-branes wrapped on $\pi_\a$ and their corresponding orientifold images on $\pi_\a^*$, and such that the RR tadpole condition (\ref{RRtad}) is satisfied. We will find a massless $U(1)_X$ for each linear  combination
\be
\pi_X =\sum_{\a=1}^K n_{X\a}\, N_\a \pi_\a,	\quad 	\quad n_{X\a} \in \IZ
\label{linX}
\ee
such that $[\pi_X] - [\pi_X^*]$ is trivial in $H_3(\CM_6,\IR)$.\footnote{If $[\pi_X] - [\pi_X^*]$ is trivial in $H_3(\CM_6,\IR)$ but not in $H_3(\CM_6,\IZ)$ then for this $U(1)$ to be massless it must have a component of RR $U(1)$ \cite{cim11}. Throughout this section we assume that Tor $H_3(\CM_6,\IZ)=0$ so that this possibility is not realized, but we will consider it again in section \ref{FGUTs}.}  Thus, the number of massless $U(1)$'s is given by $K-r$ where $r$ is dimension of the vector subspace generated by $[\pi_\a] - [\pi_\a^*]$ within $H_3^-(\mathcal M_6,\mathbb Z)$. In order to fix the normalization we pick a basis of $U(1)$'s given by
\be
\hat U(1)_\a= \frac{1}{L_\a} \sum_{\b=1}^Kn_{\a\b}\,U(1)_\b
\label{normal}
\ee
with $n_{\a\b}\in\mathbb Z$ such that $\text{g.c.d}(n_{\a1},n_{\a2},\dots,n_{lK})=1$ for all $\a=1,\dots, K$ and orthogonal, namely $n_{\a\g}n_{\g\b}=L_\a\delta_{\a\b}$. 

For a massless $U(1)_X$ we can associate the formal linear combination of 3-cycles (\ref{linX}), and we know that  there exists a 4-chain $\Sigma_X$ such that $\partial \Sigma_X=\pi_X - \pi_X^*$. Wrapping a D4-brane on $\Sigma_X$ corresponds to considering a 4d magnetic monopole of $\hat U(1)_X$ which, due to the normalization (\ref{normal}), has magnetic charge $n_X^m = 1$. Dimensionally reducing the CS action for such monopole we will find its charges with respect to the closed string $U(1)$'s from where we can read off the kinetic mixing, namely
\be\label{kinf}
f_{iX}\, =\, \frac{1}{2\,l_s^4} \int_{\Sigma_X}(J-i\F^{D4})\wedge \omega_i
\ee
where the integral is evaluated in the covering space. This expression is slightly subtle in the sense that it may depend on some discrete choices related to the pair $(\Sigma_X, \CF^{D4})$. Such subtleties can be easily removed after a proper understanding of the space of monopoles of the compactification, as we discuss in the following. 

\subsection{Monopoles and relative homology}
\label{monopole}

Besides the general formula (\ref{kinf}) for the gauge kinetic mixing between open and closed string $U(1)$'s, the previous discussion gives us an overall picture of the set of monopoles that appear in type IIA compactifications with D6-branes. On the one hand, monopoles charged under closed string $U(1)$'s are classified by D4-branes wrapping orientifold-odd 4-cycles, or in other words by the homology group $H_4^-(\CM_6,\IZ)$. On the other hand, open string $U(1)$ monopoles are classified by D4-branes wrapping odd 4-chains $\Sigma_X$ ending on the D6-branes 3-cycles $\pi_\a$ and their orientifold images $\pi_\a^*$, whose formal union we will denote as $\pi_{D6}$. The appropriate homology group that classifies such 4-chains is the relative homology group $H_4^-(\CM_6, \pi_{D6}, \IZ)$, which includes $H_4^-(\CM_6,\IZ)$ as a subgroup. In fact, we can identify $H_4^-(\CM_6, \pi_{D6}, \IZ)$ as the lattice of integral $U(1)$ magnetic charges, that contains not only monopoles charged under open string $U(1)$'s and closed string $U(1)$'s, but also bound states of those. 

Indeed, notice that the formula (\ref{kinf}) is slightly ambiguous, in the sense that we can have two different 4-chains $\Sigma_X$ and $\Sigma_X'$ with the same boundary, and so the expression for the rhs integral could be different for $\Sigma_X$ and $\Sigma_X'$. Let us temporarily simplify this formula by setting $\CF^{D4} = B$. Then, if these two chains differ by a trivial 4-cycle (that is if they belong to the same class of $H_4^-(\CM_6, \pi_{D6}, \IZ)$) then we have that $\int_{\Sigma_X} J_c \wedge \omega_i = \int_{\Sigma_X'} J_c \wedge \omega_i$ and so we get the same result for the rhs of (\ref{kinf}) independently of which chain we choose. If on the other hand $\Sigma_X$ and $\Sigma_X'$ differ by a 4-cycle $\Lambda_4^j$ such that $[\Lambda_4^j] \in H_4^-(\CM_6,\IZ)$, then the two integrals will differ by
\be
-\frac{i}{2l_s^4}\int_{\Lam_4^j} J_c \wedge \omega_i\, =\, -\frac{i}{2l_s^4}\int_{\CM_6} J_c \wedge \omega_i\wedge \omega_j \, =\,  f_{ij}
\label{clclmix2}
\ee
where $\omega_j$ is Poincar\'e dual to $\Lam_4^j$ and represents a closed string $U(1)_j$, and $f_{ij}$ is the kinetic mixing (\ref{clclmix}) between $U(1)_i$ and $U(1)_j$. The correct way to interprets this fact is that, if a D4-brane wrapping $\Sigma_X$ corresponds to a 4d monopole with unit charge under $U(1)_X$, then a D4-brane wrapping $\Sigma_X'$ has unit charge under $U(1)_X$ but also under $U(1)_j$, and so it is equivalent to a bound state of open and closed string $U(1)$ monopoles. Therefore, via the Witten effect it will obtain a electric and magnetic charge under $U(1)_i$ which is not given by the kinetic mixing $f_{iX}$, but rather by the sum of mixings $f_{iX} + f_{ij}$, see eq.(\ref{chargel}). In general, it is easy to see that the integral $\int_{\Sigma_X} J_c \wedge \omega_i$ will only depend on the homology class $[\Sigma_X] \in H_4^-(\CM_6, \pi_{D6}, \IZ)$ which as stated before is nothing but the lattice of integral $U(1)$ magnetic charges of the 4d effective theory. Hence, in order to properly use eq.(\ref{kinf}) we first need to take a basis for $H_4^-(\CM_6, \pi_{D6}, \IZ)$ and identify those 4d monopoles that have unit charge under $U(1)_X$ but no integer charge under the closed string $U(1)$'s, and then apply eq.(\ref{kinf}) with a 4-chain $\Sigma_X$ in the corresponding relative homology class. 

Let us now restore the full dependence of $\CF^{D4}$ in (\ref{kinf}) and let us see which further source of ambiguity that gives. Even if we keep $\Sigma_X$ within the same relative homology class there are infinite discrete choices of $F_2^{D4}$ such that the appropriate boundary conditions
\be
\int_{\Sig_X} F_2^{D4} \wedge \omega_i \,=\, \int_{\Sig_X} \tilde F_2^{X} \wedge \omega_i\, \equiv\,  \frac{1}{L_X} \sum_{\b=1}^K n_{X\b}\,\int_{\pi_\b} A_1^\b \wedge \omega_i
\label{WL}
\ee
are satisfied. Indeed, let us consider the case where the 4-chain $\Sigma_X$ contains a non-trivial 2-cycle $\Lam_2^j$ such that $[\Lam_2^j]$ is also non-trivial in $H_2^+(\CM_6, \IZ)$. By Poincar\'e duality on $\Sigma_X$, one may then consider a 2-form $F_2^j$ on $\Sigma_X$ which is the curvature of a vanishing line bundle on $\p \Sigma_X$ and satisfying
\be
\int_{\Sigma_X} F_2^j \wedge \g \, =\, \int_{\Lam_2^j} \g
\ee
for any closed 2-form $\g$ on $\CM_6$. Then it is easy to see that if one takes $F_2^{D4} = [F_2^{D4}]_0 + n F_2^j$ with $[F_2^{D4}]_0$ satisfying  (\ref{WL}), eq.(\ref{ch}) reads
\be
Q_i^E\, =\,\frac{1}{4\pi l_s^2}\int_{\Sig_X} \tilde F_2^{X} \wedge \omega_i + n\, \d^{ij}
\label{boundme}
\ee
where we have assumed that $[\Lam_2^j]$ is Poincar\'e dual to $[\tilde \omega^j]$ and so $\int_{\Lam_2^j} \omega_i = \d^{ij}$. This result is easily interpreted as the fact that the piece of flux $n\, F_2^j$ induces a the charge of $n$ D2-branes wrapping $\Lam_2^j$ on the D4-brane on $\Sigma_X$, so this D4-brane is actually a 4d particle with unit magnetic charge under $\hat U(1)_X$ and electric charge $n$ under $U(1)_j$. Therefore comparing (\ref{chargel}) and (\ref{boundme}) one concludes that ${\rm Im}\, f_{iX} = - \frac{1}{2l_s^4}\int_{\Sig_X} \left (B +\frac{l_s^2}{2\pi} \tilde F_2^X \right ) \wedge \omega_i$ and that in eq.(\ref{kinf}) $F_2^{D4}$ must not induce any non-trivial D2-brane charge. 

To summarize, we find that the set of monopoles in a type IIA orientifold compactification is classified by the relative homology group $H_4^-(\CM_6, \pi_{D6}, \IZ)$, where $\pi_6$ is the formal sum of the D6-brane locations. The dimension of this lattice is the total number of massless $U(1)$'s, open and closed, of the compactification, and so in some sense the space of 4d $U(1)$'s should also be classified by this same relative homology group. This is rather natural if we interpret the whole discussion above from the viewpoint of M-theory. Indeed, lifting the type IIA compactification to M-theory in a 7-dimensional manifold $\CM_7$ we have that $H_4^-(\CM_6, \pi_{D6}, \IZ)$ lifts to the homology group $H_5(\CM_7, \IZ)$, and that the $U(1)$ magnetic monopoles become M5-branes wrapping non-trivial 5-cycles in $\CM_7$ (see appendix \ref{ap:Taub} for a simple example of this lift). The $U(1)$'s themselves are classified by harmonic 2-forms in $\CM_7$, hence (assuming no torsion in homology) by the Poincar\'e dual group $H^2(\CM_7, \IZ)$. Finally, in M-theory the kinetic mixing between $U(1)$'s is given by the simple formula
\be
f_{\a\b}\, =\, -\frac{2\pi i}{l_M^9} M^I \int_{\CM_7} \phi_I  \wedge \om_\a \wedge \om_b
\label{kinM}
\ee
where $l_M$ is the M-theory characteristic length and $\phi_I$, $I = 1, \dots, b_3(\CM_7)$ runs over the harmonic 3-forms of $\CM_7$, and $M^I$ are the complex moduli associated to them. Following \cite{cim11}, from this formula one can reproduce not only the gauge kinetic mixing between type IIA closed string $U(1)$'s (\ref{clclmix}), but also the mixing between open and closed string $U(1)$'s (\ref{kinf}). In the following we will make this last connection more precise, by characterizing open string $U(1)$'s by 2-forms on $\CM_6$, that instead of representatives of $H^2(\CM_6, \IZ)$ belong to the cohomology $H^2(\CM_6-\pi_{D6}, \IZ)$, related by Lefschetz duality to the group $H_4(\CM_6, \pi_{D6}, \IZ)$ classifying the monopoles.

\subsection{Open-closed $U(1)$ mixing and linear equivalence}
\label{ociia}

The concept of linear equivalence is usually formulated to relate different $p$-cycles $\pi_p$ of a $d$ dimensional manifold $\CM_d$, being stronger than equivalence in homology. While typically one applies this concept to divisor submanifolds of a complex manifold, one may extend such definition to more general cases following \cite{Hitchin99} or the discussion in Appendix \ref{Leq}. 

Indeed, let us consider two $p$-cycles $\pi_p^a$ and $\pi_p^b$ that live in the same homology class of $\CM_d$. One can then write down the differential equation
\be
d\varpi^{(a-b)}\, =\, \d_{d-p}(\pi_p^a) - \d_{d-p}(\pi_p^b) 
\label{backrG}
\ee
where $\d_{p-d}(\pi_p^\a)$ is a bump $(d-p)$-form localized on top of the $p$-cycle $\pi_p^\a$ and transverse to it. Because $[\pi_p^a] = [\pi_p^b]$ we know that $\varpi$ is globally well-defined $(d-p-1)$-form. While there are in principle many solutions to this equation, one may in addition require that
\be
d^*\varpi^{(a-b)} \, =\, 0 \quad  \quad {\rm and} \quad \quad \int_{\Lam_{d-p-1}} \varpi^{(a-b)} \in \IZ
\label{condG}
\ee
which fixes $\varpi$ up to an harmonic representative of the cohomology group $H^{d-p-1}(\CM_d, \IZ)$. From a mathematical viewpoint, this allows to identify $\varpi$ as the connection of a gerbe.\footnote{One may actually drop the condition $d^*\varpi^{(a-b)} = 0$, which amounts to take a harmonic representative of the group $H^{d-p-1}(\CM_d-\{ \pi_p^a \cup\pi_p^b\}, \IZ)$. We will however maintain it as it simplifies the discussion.} From a physical viewpoint we will see that they are natural conditions when we want to relate $\varpi$ with an open string $U(1)$. 

Given (\ref{backrG}) and (\ref{condG}), it is easy to see that $\varpi$ admits the following global Hodge decomposition 
\be
\varpi^{(a-b)} \, =\, \omega + d^* H 
\label{Hodge1}
\ee
where $\omega$ is a harmonic $(d-p-1)$-form and $H_{d-p}$ is a globally well-defined $(d-p)$ form. We then say that the two $p$-cycles $\pi_p^a$ and $\pi_p^b$ are {\em linearly equivalent} if $[\omega] \in H^{d-p-1}(\CM_6, \IZ)$, or in other words if the two components of $\varpi$ are separately quantized. 

While the above definition is rather abstract, one may detect linear equivalence in a rather simple way as follows. Given the two $p$-cycles $\pi_p^a$ and $\pi_p^b$ let us construct a $(p+1)$-chain $\Sigma^{(a-b)}$ such that $\p \Sigma^{(a-b)} = \pi_p^a - \pi_p^b$. Then, from the discussion in  Appendix \ref{Leq} one can see that
\be
\int_{\Sigma^{(a-b)}} \tilde{\omega}_{p+1}\, =\, \int_{\CM_6} \tilde{\omega}_{p+1} \wedge \varpi^{(a-b)}\quad {\rm mod\ } \IZ
\label{chaineq}
\ee
for any closed $(p+1)$-form $\tilde{\omega}_{p+1}$ with integer cohomology class $[\tilde{\omega}_{p+1}] \in H^{p+1}(\CM_6, \IZ)$. Moreover, if $\tilde{\omega}_{p+1}$ is harmonic we have that we can replace $\varpi^{(a-b)} \raw \omega$ in the rhs of (\ref{chaineq}). Hence if we take $\tilde{\omega}_{p+1}$ to be harmonic and with integer homology class we have that $\pi_p^a$ and $\pi_p^b$ are linearly equivalent if and only if
\be
\int_{\Sigma^{(a-b)}} \tilde{\omega}_{p+1}\, \in \, \IZ \quad \quad \forall\, \Sigma^{(a-b)}\ {\rm such\ that\ } \p \Sigma^{(a-b)} = \pi_p^a - \pi_p^b
\label{Leq2}
\ee

Actually, this criterion for linear equivalence can be refined if we restrict the class of chains that enter into eq.(\ref{Leq2}), and such refinement will allow to relate the above definitions with the computation of open-closed $U(1)$ kinetic mixing. For concreteness, let us consider a simple case of interest discussed in the previous sections. Namely, we consider two D6-branes wrapping two homologous 3-cycles $\pi_3^a$ and $\pi_3^b$ of $\CM_6$. One can then write the differential equation
\be
d\varpi_2^{(a-b)}\, =\, \d_3(\pi_3^a) - \d_3(\pi_3^b)
\ee
which is nothing but (\ref{backrG}) for the particular case $d=6$, $p=3$. Requiring that $\varpi_2$ is co-closed and quantized as in (\ref{condG}) one obtains
\be
\varpi_2^{(a-b)}\, =\, c^j \omega_j + d^*H
\label{Hodge2}
\ee
where $\{\omega_j\}$ is a basis of harmonic 2-forms with integer cohomology class and $c^j \in \IR$. This definition of $\varpi_2$ only fixes the value of $c^j$ mod $\IZ$, and so one can always define $\varpi_2$ such that $c^j \in [0,1)$, $\forall \, j$. With this choice there is a 4-chain $\Sigma^{(a-b)}$ such that 
\be
\int_{\Sigma^{(a-b)}} \tilde{\omega}\, =\, \int_{\CM_6} \tilde{\omega} \wedge \varpi^{(a-b)}_2
\label{chaineq2}
\ee
for any closed 4-form $\tilde{\omega}$, and without the need of the mod $\IZ$ that appears in eq.(\ref{chaineq}).

We can also see (\ref{chaineq2}) as a consequence of Lefschetz duality between the groups $H_4(\CM_6, \pi_3^a \cup \pi_3^b, \IZ)$ and $H^2(\CM_6 - \{\pi_3^a \cup \pi_3^b\}, \IZ)$. Indeed, the 2-forms (\ref{Hodge2}) are harmonic representatives of the cohomology group $H^2(\CM_6 -\{ \pi_3^a \cup\pi_3^b\}, \IZ)$, which contains $H^2(\CM_6, \IZ)$. Changing the value of the coefficients $c^j$ by an integer number amounts to change the cohomology class by an element of $H^2(\CM_6, \IZ)$,  so choosing $c^j \in [0,1)$, $\forall \, j$ means choosing a particular class in $H^2(\CM_6 - \{\pi_3^a \cup \pi_3^b\}, \IZ)$. Then, restricting the 4-chain $\Sigma^{a-b}$ to the dual class in $H_4(\CM_6, \pi_3^a \cup\pi_3^b, \IZ)$ allows to write down (\ref{chaineq2}) without any mod $\IZ$ ambiguity.

As we saw when discussing monopoles, restricting the 4-chain $\Sigma$ to a particular relative homology class is also needed when computing the open-closed kinetic mixing from the chain integral (\ref{kinf}). In fact one can see that, if we ignore the contribution of the Wilson lines, in the present case such chain integral reads
\be
-\frac{i}{2l_s^4} \int_{\Sigma^{(a-b)}}  J_c \wedge \omega_i \,= \, -\frac{i}{2l_s^4} \int_{\CM_6} J_c \wedge \omega_i \wedge \varpi_2^{(a-b)}\, =\, -\frac{i}{2}  \CK_{ij\hat{k}} T^{\hat{k}} c^j \, =\, f_{ij}c^j
\ee
Finally, for $c^j \in [0,1)$ linear equivalence between $\pi_3^a$ and $\pi_3^b$ amounts to require that $c^j =0$, $\forall\, j$. That is, the equivalence $\pi_3^a \sim \pi_3^b$ corresponds to the vanishing of the kinetic mixing $f_{i(a-b)}$ of $\oh[U(1)_a-U(1)_b]$ with any $U(1)_i$ from the closed string sector. 

All this discussion can be easily generalized for the case where we have more than two D6-branes wrapping 3-cycles the same homology class. For each massless $U(1)_X$ of the open string sector given by a linear combination of 3-cycles (\ref{linX}) such that $[\pi_X] =0$ we can define the 2-form $\varpi_X$ by
\be
d\varpi_X\, =\, \sum_\a n_{X\a} N_\a \d_3 (\pi_\a)
\label{fakeX}
\ee
and such that $\varpi_2$ co-closed and has integer relative homology class. This fixes $\varpi_X$ to be of the form (\ref{Hodge2}), where again we impose that $c^j \in [0,1)$. Then we find that the kinetic mixing between $U(1)_X$ and a $U(1)_i$ of the closed string sector is given by
\be
f_{iX}\, =\,  -\frac{i}{2l_s^4}  \int_{\CM_6}  J_c  \wedge \omega_i \wedge \varpi_{X}
\label{kinXfin}
\ee
which vanishes if the combination $\pi_X$ is linearly trivial, since then $c^j=0$, $\forall\, j$. Notice that (\ref{kinXfin}) reduces to (\ref{kinf}) if we replace $\int_{\CM_6} \varpi_X \wedge\, \raw\, \int_{\Sigma_X}$, and then take $\CF^{D4} = B|_{\Sigma_X}$ to ignore the Wilson line dependence. Finally, one can generalize this expression to the case of orientifold compactifications by modifying (\ref{fakeX}) in the obvious way and by taking $\varpi_X$  with the appropriate orientifold parity, or more precisely by taking into account that $[\varpi_X] \in H^2_+(\CM_6-\pi_{D6}, \IZ)$. 

To summarize, in the absence of Wilson lines, the vanishing of the gauge kinetic function $f_{iX}$ that mixes open and closed string $U(1)$'s corresponds to the linear equivalence of the two or more cycles that define the open string $U(1)$. The actual expression for this gauge kinetic mixing can either be expressed via a chain integral as in (\ref{kinf}) or as an integral all over the compactification manifold as in (\ref{kinXfin}). The later is suggestive in the sense that it resembles the well-known expression for the closed-closed kinetic mixing (\ref{clclmix}), with the difference that now the open string $U(1)$ is represented by $\varpi_X$, which is a harmonic representative of $H^2_+(\CM_6-\pi_{D6}, \IZ)$. 

As we have seen, the equality between linear equivalence and vanishing kinetic mixing is no longer true when we take the Wilson line dependence into account, and one needs to define a generalized notion of linear equivalence. Nevertheless, we will see that even in this case one is able to express the open-closed gauge kinetic mixing as an integral of the form (\ref{kinXfin}) over the whole compactification manifold. In order to do that, however, we first need to discuss the role played by relative cohomology groups in the computation of gauge kinetic functions.

\subsection{Wilson lines and relative cohomology}

 Generically, a D6-brane Wilson lines correspond to harmonic one-forms $\zeta$ on a 3-cycle $\pi$ whose one-cycles are trivial in the homology of the ambient space $\CM_6$. Because of that, we cannot relate $\zeta$ to any closed one-form of $\CM_6$. Nevertheless, the Poincar\'e dual 2-cycle $\rho$ of $\zeta$ in $\pi$ can be nontrivial in $\CM_6$ and, if this is the case, such Wilson line will enter in the gauge kinetic functions $f_{iX}$ mixing open and closed string $U(1)$'s. One way to detect this is by looking at the field strength $\CF^{D4}$ of the appropriate open string monopole, which will depend on this Wilson line and so will the kinetic mixing via (\ref{kinf}). Physically, due to the Wilson line $\zeta$ one must induce a D2-brane charge along $\rho$ on the D4-brane which is the 4d open string monopole, and this translates into a open-closed $\theta$ angle via the Witten effect.

The appropriate way to incorporate this effect into a formula of the form (\ref{kinXfin}) is via the use of relative cohomology groups. Let in particular look at the relative cohomology $H^n(\CM_6,\pi_{D6})$, where for illustrative purposes we can take $\pi_{D6} = \pi_3^a \cup \pi_3^b$ as the union of two homologous 3-cycles. The groups $H^n(\CM_6,\pi_{D6})$ are obtained by considering pairs of forms such that
\be
(\alpha_{n},\beta_{n-1})\quad \in \quad \Omega^{n}(\CM_6) \times \Omega^{n-1}(\pi_{D6}) 
\label{relpair}
\ee
and constructing the cohomology in the standard way with the differential
\be
d(\alpha_{n},\beta_{n-1})\, =\, (d\a_n, \iota_{\pi_{D6}}^*(\a_n) - d\beta_{n-1})
\label{reldif}
\ee
It is then easy to see that any element of the form $(0,\beta_{n-1})$, with $\beta_{n-1}$ a non-trivial closed form in $\pi_{D6}$, is also nontrivial in $H^n(\CM_6,\pi_{D6},\IZ)$ if it cannot be written as the pull-back $\iota_{\pi_{D6}}^*(\g_{n-1})$ of a globally well-defined closed form $\g_{n-1}$ of $\CM_6$. For instance, for $\pi_{D6} = \pi_3^a \cup \pi_3^b$ with $\pi_3^a$, $\pi_3^b$ homotopic 3-cycles with a harmonic form $\beta_{n-1}^{a,b}$ on each of them, a choice such that $\b_{n-1}^a \neq \b_{n-1}^b$ will always correspond to a non-trivial element of $H^n(\CM_6,\pi_{D6})$, because there is no closed form $\g_{n-1}$ of $\CM_6$ that would give different integrals in the two corresponding $(n-1)$-cycles. In particular, we have that Wilson lines $\theta_a\zeta$ and $\theta_b\zeta$ with $\theta_a \neq \theta_b$ corresponds to a non-trivial element of $H^2(\CM_6,\pi_{D6})$. Moreover, by using the definition (\ref{reldif}) one can see that this Wilson line configuration is equivalent to $(\Upsilon_2,0)$ where $\Upsilon_2 =d \Theta_1$ is an exact 2-form of $\CM_6$ such that $\Theta_1|_{\pi_3^\a} = \th_\a\zeta$, $\a=a,b$. 
Finally, it is easy to generalize this construction to the case that there is more than two D6-branes and several Wilson lines on each of them, the final result being that for each open string $U(1)_X$ there is a bulk 2-form $\Upsilon_2^X = d \Theta_1^X$ such that $\Theta_1^X$ restricts (up to an exact form) to the corresponding Wilson line on each of the 3-cycles $\pi_3^\a$ that enter into the linear combination (\ref{linX}).

Given this setup, it is easy to see that the Wilson line contribution to the gauge kinetic mixing can be written as
\be
-\frac{i}{4\pi l_s^2} \int_{\CM_6} \Upsilon_2^X \wedge \omega_i \wedge \varpi_X =  \frac{i}{4\pi l_s^2} \int_{\CM_6} \Theta_1^X \wedge \omega_i \wedge d \varpi_X = -\frac{i}{4\pi l_s^2} \sum_\a n_{X\a} N_\a \int_{\pi_\a} A_1^\a \wedge \omega_i
\ee
replacing the Wilson line integral (\ref{wldef}) that defines $\tilde F_2^{X}$, by an integral over the whole manifold $\CM_6$ involving the 2-form $\Upsilon_2^X$. One possible way to interpret this is via the trading of the Wilson line background by a shift in the B-field background, as done in \cite{dps14}. As a result, the total gauge kinetic mixing can be expressed via the equation (\ref{kinXfin}) with the replacement $J_c \raw J_c + \frac{l_s^2}{2\pi} \Upsilon_2^X$.

\subsection{Summary}

In this section we have computed the kinetic mixing between open and closed string $U(1)$'s in type IIA compactifications. Via the Witten effect we have obtained the  expression
\be\label{kinsum}
f_{iX}\, =\, \frac{1}{2\,l_s^4} \int_{\Sigma_X}(J-i\F^{D4})\wedge \omega_i
\ee
for the mixing between a closed $U(1)_i$ and a open string $U(1)_X$. Here $[\omega_i] \in H^2_+(\CM_6, \IZ)$ and $(\Sigma_X,\F^{D4})$ is a 4-chain and with gauge bundle on it, describing a open string $U(1)$ monopole made up of a D4-brane connecting the D6-branes of the compactification. Notice that this expression depends on very few, local data of the compactification, as opposed to the threshold corrections that induce a mixing between open string $U(1)$'s.

As a byproduct of our discussion we observed that, in the presence of massless open string $U(1)$'s, the lattice of 4d monopoles of a type IIA compactification is no longer given by $H_4^-(\CM_6, \IZ)$, but rather by the relative homology group $H_4^-(\CM_6, \pi_{D6}, \IZ)$. This group enlarges $H_4^-(\CM_6, \IZ)$ by adding classes of 4-chains ending on combination of 3-cycles $\pi_{D6} = \cup_\a \pi_3^\a$ wrapped by D6-branes of the compactification. Replacing $H_4^-(\CM_6, \IZ)$ by $H_4^-(\CM_6, \pi_{D6}, \IZ)$ is related, by Lefschetz duality, to  $H^2_+(\CM_6, \IZ) \raw H^2_+(\CM_6-\pi_{D6}, \IZ)$, which adds to the closed 2-forms of $\CM_6$ those 2-forms that are closed only up to the D6-brane locations $\pi_{D6}$. Examples of these new forms are the 2-forms $\varpi_X$, defined by (\ref{fakeX}) and used to detect linear equivalence between two or more 3-cycles. These new 2-forms can be used to compute open-closed kinetic mixing via the following expression
\be
f_{iX}\, =\,  \frac{1}{2l_s^4}  \int_{\CM_6}  (J - i \CF) \wedge \omega_i \wedge \varpi_{X}
\label{kinXfinWL}
\ee
where $[\varpi_X]$ is a new element of the extension $H^2_+(\CM_6-\pi_{D6}, \IZ)$ that represents $U(1)_X$, and $\CF = B + \frac{l_s^2}{2\pi} \Upsilon_2^X$ where $[ \Upsilon_2^X] \in H^2_-(\CM_6,\pi_{D6})$ is defined as in the last section. Finally, (\ref{kinXfinWL}) reduces to (\ref{kinsum}) if we replace $\int_{\CM_6} \varpi_X \wedge\, \raw\, \int_{\Sigma_X}$, and take $\CF^{D4} = \CF|_{\Sigma_X}$. 

Eq. (\ref{kinXfinWL}) is quite suggestive from the viewpoint of M-theory, since by replacing 
\begin{equation}
\begin{array}{rcl}
\CM_6 & \raw & \hat{\CM}_7 \\
J_c +F_2 & \raw & M^I \phi_I \\
\varpi_X & \raw & \omega_X
\end{array}
\end{equation}
with $\omega_X$ an harmonic 2-form on $\hat{\CM}_7$ one recovers the M-theory expression (\ref{kinM}) for the kinetic mixing of two $U(1)$'s. All these replacements are standard when lifting a type IIA compactification to M-theory except perhaps the last one, which suggests that a harmonic representative of $H^2_+(\CM_6 - \pi_{D6}, \IZ)$ is related to an harmonic representative of $H^2(\hat\CM_7, \IZ)$. This is related to the results of section \ref{monopole}, where in order to match monopole lattices we concluded that $H_4^-(\CM_6, \pi, \IZ)$ should be identified with $H_5(\hat\CM_7, \IZ)$. A simple example of this connection is given in appendix \ref{ap:Taub}, while a general analysis is left for future work.

\section{Linear equivalence of magnetized D-branes}
\label{IIBmix}

In our analysis of type IIA compactifications we have found that the kinetic mixing between open and closed string $U(1)$'s can be computed by means of a simple chain formula, whose physical meaning can be understood as the Witten effect applied to D-brane $U(1)$ monopoles. Moreover, we found that just like the set of monopoles is classified by a relative homology group, the set of U(1)'s is classified by a dual cohomology group. In particular, one is able to characterize the massless open string $U(1)$'s in terms of a pair of bulk 2-forms $(\varpi, \Upsilon_2)$ from which one can reproduce the chain formula for the kinetic mixing. Up to Wilson line contributions, this kinetic mixing will vanish when the set of D6-branes defining the $U(1)$ are linearly equivalent, or in other words when $\varpi$ is orthogonal to any harmonic 2-form in the compactification manifold. 

In this section we would like to extend this picture to type IIB compactifications. The novelty of these compactifications is that they contain magnetized branes, and so the objects to consider are bound state of 3, 5 and 7-branes. However, just like in the type IIA case, we will be able to describe open-closed kinetic mixing by using the Witten effect, and to characterize open string $U(1)$'s in terms of bulk forms $(\varpi, \U)$. Also, the fact that the open-closed mixing vanishes can be translated into a version of linear equivalence adapted to D-brane bound states. Finally, we will derive a supergravity-like formula to compute kinetic mixing between open and closed string $U(1)$'s, and apply it to F-theory GUT models with hypercharge flux breaking \cite{Beasley:2008kw,dw08}.

\subsection{Type IIB orientifolds with O3/O7-planes}

Let us then consider type IIB string theory compactified on $\IR^{1,3} \times \CM_6$, with $\CM_6$ a Calabi-Yau manifold, and mod it out by the orientifold action $\Omega_p (-1)^{F_L} \sigma$, where now $\sigma$ is a holomorphic involution $\CM_6$ such that
\begin{equation}
\sigma J=J\ , \qquad \sigma\Omega=-\Omega
\end{equation}
The fixed loci of $\sigma$ are points and/or complex 4-cycles of $\CM_6$, where O3 and O7-planes are respectively located. The cancellation of RR tadpoles imposes that
\be
\sum_\a N_\a([\pi_\a]+[\pi_\a^*])-8[\pi_O]=0
\label{RRtadD7}
\ee
where $\pi_\a$ are the complex 4-cycles wrapped by D7-branes, $\pi_\a^*$ are their orientifold images, and $\pi_O$ the divisors wrapped by the orientifold planes. Besides (\ref{RRtadD7}) one needs to impose that the total D5 and D3-brane charges of the compactification vanish. 

Dimensionally reducing the 10d type IIB supergravity action one encounters a series of massless fields that arise from the closed string sector of the theory. As before, these are classified by the harmonic forms of $\CM_6$ with a definite parity under the action of $\sig$. 
We now take the following basis of harmonic forms with integer cohomology class
\be\nonumber
\begin{array}{ccc}
&\sigma\rm{-even}&\sigma\rm{-odd}\\
\rm {2-forms}\quad&\omega_i\quad i=1,\dots,h_+^{1,1}\quad&\omega_{\hat i}\quad \hat i=1,\dots,h_-^{1,1}\\
\rm {3-forms}\quad&\a_I\quad I=0,\dots, h^{1,2}_+ \quad& \tilde \a_{\hat I}\quad {\hat I}=0,\dots, h^{1,2}_- \\
& \b^I\quad I=0,\dots, h^{1,2}_+ \quad& \tilde\b^{\hat I}\quad {\hat I}=0,\dots, h^{1,2}_- \\
\rm {4-forms}\quad&\tilde\omega^{i}\quad i=1,\dots,h_+^{1,1}\quad&\tilde\omega^{\hat i}\quad {\hat i}=1,\dots,h_-^{1,1}\end{array}
\ee
and with normalization
\be
\int_{\mathcal{M}_6}\omega_i\,\w\,\tilde \omega^j=l_s^6\d_i^j,\quad\int_{\mathcal{M}_6}\omega_{\hat i}\,\w\,\tilde \omega^{\hat j}=l_s^6\d_{\hat i}^{\hat j},\quad\int_{\mathcal{M}_6}\a_I\,\w\,\b^J=l_s^6\d_I^J, \quad \int_{\mathcal{M}_6}\tilde \a_{\hat I}\,\w\,\tilde \b^{\hat J}=l_s^6\d_{\hat I}^{\hat J}
\label{intersectionB}
\ee
The next step is to expand the RR 4-form potential $C_4$ in the $\sigma$-even harmonic forms $\om_i$, $\a_I$, $\b^I$, $\tilde{\om}^i$. Since the field strength of $C_4$ must satisfy the 10d self-duality condition $\hat F_5=*_{10}\hat F_5$, it is convenient to define the following basis of complex harmonic 3-forms
\be
\g_I \, =\, \a_I + i f_{IJ} \beta^J
\label{defg}
\ee
where $f_{IJ}$ are function of the complex structure moduli chosen so that $\g_I$ is a $(2,1)$-form. Then it is easy to see that if we expand the 4-form potential as
\begin{equation}
C_4=\sum_I(A_1^I\wedge \re\, \g_I-V_1^I\wedge \im \, \g_I)+\sum_i \left(C_2^i\wedge \omega_i-\textrm{Re}(T^i)\tilde{\omega}^i\right)
\end{equation}
then 10d self-duality of $\hat F_5$ implies that $*_4dA^I_1 =  dV^I_1$. We then obtain $h^{1,2}_+$ vector multiplets $A_1^I$ and $h_+^{1,1}$ axions $\re(T^i)$, the other 4d modes being dual degrees of freedom. Finally, using (\ref{intersectionB}) it is easy to check that $f_{IJ}$ is precisely the gauge kinetic function for the closed string $U(1)$'s \cite{deWit:1984px,Andrianopoli:1996cm,Grimm:2004uq}. Notice for instance that a D3-brane wrapping a 3-cycle in the Poincar\'e dual class of $\a_J$ will not only have magnetic charge under the closed string $U(1)$ generated by $A_1^J$, but also an electric charge under $A_1^I$ proportional to $\im \, f_{IJ}$, as expected from the Witten effect. 

Besides above spectrum there will be 4d massless fields arising from the open string sector of the compactification. In particular, the sector of a single D7-brane wrapping a holomorphic 4-cycle $S_\a$ of $\CM_6$ is given by
\bea
\label{WLD7}
A_1^{\a}&=&\frac{1}{l_s}(\pi A_1^{4d,\a}+ a_\a^j\,A_j + c.c. )\\
\phi_\a&=& \Phi_\a + c.c.\, =\, \Phi_\a^m X_m + c.c.
\label{posD7}
\eea
where $A_j$ are a basis of $H^{(0,1)}(S_\a)$ and $X_m$ are holomorphic sections of the normal bundle of $S_\a$, which are in one to one correspondence with the homology group $H^{(2,0)}(S_\a)$ by contraction with $\Omega$. As a result, the Wilson line moduli $a_\a^j$ and the position moduli $\Phi_\a^j$ are each complex 4d scalar fields by themselves.\footnote{We normalize the fields $\Phi_\a$ as $\Phi_\a=\frac{y_\a}{l_s}$ where $y_\a$ are the transverse coordinates to the brane $\a$.} One can easily generalize this spectrum to the case of a stack of $N_\a$ D7-branes, as well as to include the effect of the orientifold projection. We refer the reader to \cite{jl04} for further details in this direction.

\subsection{Separating two D7-branes}

For simplicity let us first consider two D7-branes wrapping 4-cycles $S_a$ and $S_b$ of $\mathcal M_6$, and threaded respectively by worldwolume fluxes $\bar F_a$ and $\bar F_b$, leaving the action of the orientifold for later. Like with the D6-branes we assume that $S_\a$, $\a=a,b$ lie in the same homology class and that they can both be obtained after normal deformations $\phi_\a$ of a reference 4-cycle $S$. As before, we can reduce the CS action for these branes and read off the terms that yield St\"uckelberg masses for the open string $U(1)$'s as well as their kinetic mixing with the closed string sector. The relevant terms of the CS action now are
\be
S_{CS}^\a\supset\mu_7\int_{\mathbb R^{1,3}\times S_\a}P[C_6\,\w\,\F_2^\a+\frac{1}{2}C_4\,\w\,\F_2^\a\,\w\,\F_2^\a]
\ee
which upon dimensional reduction yield \cite{jl04}
\bea\label{redb}
S_{CS}^\a&\supset&\frac{\pi}{l_s^7}\int_{\mathbb R^{1,3}}\tilde C_2^i\,\w\,F_2^\a\int_{S_\a}\tilde\om^i+\frac{1}{2 l_s^7}\int_{\mathbb R^{1,3}} C_2^i\,\w\,F_2^\a\int_{S_\a}\bar F_2^\a\,\w\,\om^i\\\nonumber
&+&\frac{1}{2 l_s^6} \int_{\mathbb R^{1,3}}F_2^\a\, \w\, F_2^{RR,I}\int_{S_\a} \re\, (a_\a\,\w\, \g_I+ \iota_{\Phi_\a}\g_I\,\w\,\bar F_2^\a)\\\nonumber
&-&\frac{1}{2 l_s^6} \int_{\mathbb R^{1,3}}F_2^\a\, \w\, *F_2^{RR,I}\int_{S_\a} \im\, (a_\a\,\w\, \g_I+ \iota_{\Phi_\a}\g_I\,\w\,\bar F_2^\a)
\eea
where we have used
\bea
C_4&=& C_2^i\,\w\, \om^i+ A_1^I\wedge \re\, \g_I-V_1^I\wedge \im \, \g_I + \dots\\
C_6&=& \tilde C_2^i\,\w\, \tilde \omega^i+\dots
\eea
with $\om^i$, $\tilde \om^i$ running over all (1,1), (2,2)-forms of $\CM_6$. We have also used that $a_\a = a_\a^j\,A_j$ is a (0,1)-form and $\bar F_2^\a$, $\iota_{\Phi_\a} \g_I$ are (1,1)-forms of $S_\a$.  
Finally, we omitted any B-field contribution since it can be simply recovered by replacing $\frac{l_s^2}{2\pi}F\rightarrow\mathcal F$. 

The terms in the first line of (\ref{redb}) correspond to the St\"uckelberg mass for the open string $U(1)$ and the second line gives the kinetic mixing with the closed string sector. Considering both D7-branes $a$ and $b$, the combination $U(1)_a + U(1)_b$ will be massive due to the first term in (\ref{redb}). In order to keep $U(1)_{(a-b)}\equiv \oh[U(1)_a - U(1)_b]$ massless we impose that $[\bar F_2^a]=[\bar F_2^b]\equiv [\bar F_2]$. Notice that the St\"uckelberg mass for $U(1)_{(a-b)}$ is proportional to
\be
\int_{S}(\bar F_2^a-\bar F_2^b)\,\w\,\om^i
\ee
and so setting both worldvolume flux equal on $S$ prevents this $U(1)$ from getting a mass.
There is a subtlety in this statement which is important for the case of F-theory GUTs. Namely, in order to keep $U(1)_{(a-b)}$ massless the class $[\bar F_2^a-\bar F_2^b]$ should only be zero as an element of $H^2(\mathcal M_6)$ and it could be non-trivial in $H^2(S)$ \cite{Buican:2006sn}. We will assume that the fluxes are the same also in $S$ and leave the more involved case for the next section. 

A computation similar to section \ref{IIAmix} shows that the  mixing between $U(1)_{(a-b)}$ and $U(1)_I$ is
\be\label{kinbm}
f_{I(a-b)}=-\frac{i}{4\pi l_s^4}(a^j_a-a^j_b)\int_S A_j\,\w\,\g_I-\frac{i}{4\pi l_s^4}(\Phi_a^m-\Phi_b^m)\int_S\iota_{X_m}\g_I\,\w\,\bar{F}_2
\ee
Comparing this expression to (\ref{kinb}) and (\ref{kinf}) one can guess what the mixing is in the case where the branes are not homotopic, namely
\be\label{kinbb}
f_{I(a-b)}=-\frac{i}{2 l_s^5}\int_{\G}\g_I\,\w\,\tilde{\F}
\ee
where $\G$ is a 5-chain with $\p\G=S_a-S_b$ and $\tilde{\F}$ is a 2-form defined on $\Gamma$ such that $d\tilde\F=0$ and $\tilde{\F}|_{S_\a}=\bar{\F}_\a$. We implicitly include the contribution coming from the Wilson lines in the boundary $\p\G$ as in (\ref{wldef}) (see also the comment below eq.(\ref{ch})). Finally, notice that $\F$ also includes the contribution from the $B$-field, but that it vanishes in the case that $H=dB=0$, since then $B$ is an harmonic (1,1)-form and so $\g_I \wedge B \equiv 0$. 

As before, we can also arrive at (\ref{kinbb}) by computing the electric charge of a 4d $U(1)_{(a-b)}$-monopole with respect to the closed string $U(1)$'s. Indeed, consider a D5-brane wrapped on $W\times \G$ where $W$ is a worldline in $\mathbb R^{1,3}$, and with worldvolume flux $\tilde{\F}$ along $\G$. This corresponds to a monopole in 4d with unit magnetic charge under $U(1)_{(a-b)}$. The CS action for this objects has a piece of the form
\be
S_{CS}^{D5}\supset \mu_5 \int_{W\times \G} C_4\,\w\,\tilde{\F}=\mu_5\int_{\G} \tilde{\F}\,\w\,\re\, \g_I  \int_WA_1^I - \mu_5\int_{\G} \tilde{\F}\,\w\,\im\, \g_I  \int_WV_1^I
\ee
so the induced electric and magnetic charges under $U(1)_I$ are
\be\label{charge}
Q_I^E=\frac{1}{2l_s^5}\int_{\G}\re\, \g_I \,\w\, \tilde\F\quad \quad \quad \tilde Q_I^M=-\frac{1}{2 l_s^5}\int_{\G}\im\, \g_I \,\w\, \tilde\F
\ee
which reproduces (\ref{kinbb}) by virtue of the Witten effect.

The effect of the orientifold projection is also quite similar to the type IIA case. We have to take into account that $\omega^i$ and $\tilde \omega^i$ in (\ref{redb}) only run over $\sigma$-even forms of $\CM_6$, and we need to include the orientifold image for every D7-brane. This leads to
\be\label{kinbbo}
f_{I(a-b)}=-\frac{i}{2 l_s^5}\int_{\G}\g_I\,\w\,\tilde\F
\ee
with $\G$ an orientifold-odd chain such that $\p\G=S_a-S_b-S_a^*+S_b^*$. Again, the contribution from the B-field vanishes in case that $H=0$, which we will assume in the following.

\subsection{Monopoles and generalized cycles}

Let us now make a detour and  rephrase the previous example in the language of generalized complex geometry. This will allow us to easily extend the above expression for the $U(1)$ mixing to more general situations, and in particular to the case of interest in F-theory GUTs to be analyzed in section \ref{FGUTs}. In addition, this formalism properly treats D-branes with worldvolume fluxes, and so it allows to generalize the concept of linear equivalence of submanifolds \cite{Hitchin99} to bound states of D-branes, as well as to derive a supergravity-like formula for its kinetic mixing. Finally,  generalized geometry has been shown to be the right framework to include the effect of background fluxes in both type IIA and type IIB $\CN=1$ compactifications, and hence it should be the appropriate tool to understand $U(1)$ kinetic mixing in the context of moduli stabilization. 

In generalized complex geometry a D$p$-brane is a pair $(\Sigma,\F)$ where $\Sigma$ is a $(p+1)$-cycle and $\F$ is the worldvolume field strength together with the $B$-field, $\F=F+B$.\footnote{See Appendix \ref{gh} for some basic definitions of generalized submanifolds and their homology \cite{em07}.} Thus, the two D7-branes of the last section correspond to $(S_a,\F_a)$ and $(S_b,\F_b)$ and the condition to have a massless $U(1)$ is that there should be a linear combination of the two that is trivial in generalized homology. Indeed, in the previous example we had
\be\label{an}
(S_a,\F_a)-(S_b,\F_b)=\hat\p(\G,\tilde{\F})
\ee
where $\hat\p$ is the generalized boundary operator defined in Appendix \ref{gh}, while $\G$ and $\tilde \F$ are the 5-chain and worldvolume flux defined below (\ref{kinbb}). In particular, $\tilde \F$ satisfies 
\be
d\tilde\F=0,\qquad \tilde \F|_{S_a}=\F_a,\qquad \tilde\F|_{S_b}=\F_b.
\ee
where we have used that $H=0$.
In order to compute the kinetic mixing we look at the CS action of a D5-brane wrapped on $(W\times\G,\F)$ which can be written as
\be
S_{CS}=\mu_5\int_{W\times\G}C\,\w\,e^{\tilde{\F}}
\ee
where $C$ is the RR polyform. Since the closed string $U(1)$'s arise from $C_4$ it suffices to look at that term. Upon dimensional reduction it yields
\bea
S_{CS}&\supset&\mu_5\int_{\G}\re\, \g_I\,\w\,e^{\tilde{\F}}\int_WA_1^I\ -\mu_5\ \int_{\G}\im\, \g_I\,\w\,e^{\tilde{\F}}\int_WV_1^I\\ \nonumber
& =& \mu_5\,j_{(\G,\tilde{\F})}(\re\, \g_I)\int_WA_1^I\ -\mu_5\, j_{(\G,\tilde{\F})}(\im\, \g_I)\int_WV_1^I
\eea
where we have introduced the current associated to the generalized chain $(\G,\tilde{\F})$, dubbed $j_{(\G,\tilde{\F})}$, and used the fact that the D5 flux $\tilde{\F}$ is magnetic so it has no component along the time direction $W$. Thus, the $U(1)_I$ electric and magnetic charges of the D5-monopole are
\be
Q^I_E=\frac{1}{2 l_s^5}\,j_{(\G,\tilde{\F})}(\re \, \g_I) \quad \quad \quad \tilde Q^I_M=-\frac{1}{2 l_s^5}\,j_{(\G,\tilde{\F})}(\im \, \g_I)
\ee
from where we can read off the kinetic mixing. Applying this formalism to this example is somewhat cumbersome but it shows that the only relevant information for the mixing is in the so-called generalized chain $(\G,\tilde{\F})$, and that this procedure for computing the mixing can be generalized to more involved setups. 

We now consider a case which is closely related to the hypercharge $U(1)_Y$ in F-theory GUTs and for which the formalism of generalized geometry turns out to be quite useful. Suppose again that we have two D7-branes wrapped on 4-cycles $S_a$ and $S_b$, both homotopic to $S$. This implies that our D7-branes are described by the generalized cycles $(S_a,\F_a)$ and $(S_b,\F_b)$ such that $[S_a]=[S_b]\equiv [S]$. In addition we assume that $[\F_a]=[\F_b]$ in the cohomology of the total space but with $[\F_a]$ and $[\F_b]$ different in $H^2(S)$.\footnote{We are also assuming that $\int_{S_a} \CF_a \wedge \CF_a = \int_{S_b} \CF_b \wedge \CF_b$.}  In that case we get a massless $U(1)$ so there must exist a generalized chain that connects the D7-branes but it cannot be the same one as before. Indeed, suppose it were $(\G,\tilde\F)$. The chain $\G$ connecting the submanifolds is topologically $S\times I$ where $I$ is an interval with coordinate $t\in[0,1]$ so $\G$ can be sliced in 4-cycles $S_t\simeq S$. Thus, the 2-form $\tilde\F$ on $S\times I$ defines a family of 2-forms $\tilde\F_t$ on $S_t$ such that $\tilde \F_0=\F_a$ and $\tilde\F_1=\F_b$. Since $\tilde\F$ is continuous and $d\tilde\CF=0$ we have that $[\tilde\F_0]=[\tilde\F_1]$ in $H^2(S)$, contradicting our assumption.

This means that we have to look for another candidate to be the generalized chain associated to the massless $U(1)$. The strategy to find it is to 
realize that the quantized part of the worldvolume flux $\bar F_\a$ induces a D5-brane charge that can be related by Poincar\'e duality to a 2-cycle class $[\Pi_\a]$ of $H_2(S_\a,\IZ)$. If instead of two magnetized D7-branes we had two D7-branes on $S_a$, $S_b$ with $\bar F_a = \bar F_b =0$ and two D5-branes on the representative 2-cycles $\Pi_a$, $\Pi_b$, then we would have the same D-brane charges as in the magnetized system,\footnote{We are ignoring induced D3-brane charges since they become irrelevant in the orientifold case.} and it would be simple to connect them by means of a 5-chain $\G$ and a 3-chain $\Sigma$ such that $\p\Sig=\Pi_a-\Pi_b$. So the only ingredient that we need is a generalized chain that interpolates between a magnetized D7-brane and a D7+D5-brane pair.

Such generalized chain is given by the following equation 
\be
(S_a,\F_a)=(S_a,0)+(\Pi_a,0)+\hat\p\left [ -(\G_a,\tilde \F_a) + (\G_a,\tilde\F_{\Pi_a})\right ]
\ee
with $\Pi_a\subset S_a$ a 2-cycle such that $[\Pi_a]={\rm P.D.\,}[F_a]$, $\G_a$ a 5-chain with $\p \G_a=S'_a-S_a$ and 
\bea
d\tilde\F_a=0 & \tilde \F_a|_{S_a}=\bar F_a& \tilde \F_a|_{S'_a}=\bar F_a' \\
d\tilde\F_{\Pi_a}=\d^{(3)}_{\G_a}(\Pi_a)& \qquad \tilde\F_{\Pi_a}|_{S_a}=0\qquad &\tilde\F_{\Pi_a}|_{S'_a}= \bar F_a'
\eea
where for simplicity we have removed the presence of the B-field, which anyway will not appear in our final result. 
This looks rather messy but its interpretation as a physical process is simple. The term $-\hat\p(\G_a,\tilde \F_a)$ corresponds to moving the brane from $S_a$ to a reference 4-cycle $S_a'$ keeping the worldvolume flux fixed. The second term $\hat\p(\G_a,\tilde \F_{\Pi_a})$ is responsible for moving the brane back to $S_a$ and removing the flux leaving the remnant $(S_a,0)+(\Pi_a,0)$ which represents a D7 on $S_a$ and a D5 on $\Pi_a$ each of them without fluxes. The fact that we have to move the brane back and forth is a technicality that regularizes the differential equations, and eventually we will take the limit in which we do not move the brane at all (see the discussion in Appendix \ref{cal}). 

Doing the same with the brane $b$ we arrive at
\be
(S_a,\F_a)-(S_b,\F_b)=\hat\p\left [ (\G,0)+(\Sig,0)-(\G_a,\tilde \F_a)+(\G_b,\tilde \F_b) + (\G_a,\tilde\F_{\Pi_a})-(\G_b,\tilde\F_{\Pi_b})\right ]
\ee
with $\Sig$ a 3-chain such that $\p\Sig=\Pi_a-\Pi_b$. This equation describes a process in which the worldvolume fluxes are turned into D5-branes at both $S_a$ and $S_b$, and these are connected to each other by means of the 3-chain $\Sig$, while the D7s are connected via $\G$. Thus, we have found the generalized chain $(\mathfrak S,\mathfrak F)$ associated with the open string massless $U(1)$.

According to our previous discussion the kinetic mixing with the closed string $U(1)_I$ will be proportional to $j_{(\mathfrak S,\mathfrak F)}(\g_I)$. One still needs to show that the result is independent of the arbitrary choices made to define $(\mathfrak S,\mathfrak F)$. We relegate the proof to Appendix \ref{cal}, in which we also derive the following more convenient expression
\be\label{jfrak}
j_{(\mathfrak S,\mathfrak F)}(\g_I)=\int_{\Sig}\g_I+\frac{1}{2\pi l_s}\int_{S_a}\g_I\,\w\,A_{\Pi_a}-\frac{1}{2\pi l_s}\int_{S_b}\g_I\,\w\,A_{\Pi_b}.
\ee
with $d A_{\Pi_a}=2\pi l_s[\F_a-\d^2_{S_a}(\Pi_a)]$ and $d A_{\Pi_b}=2\pi l_s[\F_b-\d^2_{S_b}(\Pi_b)]$. Again, one can show that this expression is independent of the choice of $\Pi_a$ and $\Pi_b$. Notice that these equations do not fix the harmonic piece of $A_{\Pi_a}$ and $A_{\Pi_b}$. In order to match the result obtained from dimensional reduction we take them to be the Wilson lines in each D7-brane. Introducing the normalization factor we finally find
\be\label{mixfrako}
f_{I(a-b)}=-\frac{i}{2  l_s^3}\left [\int_{\Sig}\g_I+\frac{1}{2\pi l_s}\left(\int_{S_a}\g_I\,\w\,A_{\Pi_a}-\int_{S_b}\g_I\,\w\,A_{\Pi_b}\right)\right ].
\ee
In Appendix \ref{cal} we show that this expression reduces to (\ref{kinbb}) when $[\bar F_a]=[\bar F_b]\in H^2(S)$.\footnote{The dependence on $A_{\Pi_\a}$ was overlooked in the expression for the mixing in Appendix B of \cite{cim11}.}
Finally we can easily extend this analysis to the orientifold case, where  we have
\be\label{mixfrako2}
f_{I(a-b)}=-\frac{i}{4  l_s^3}\left [\int_{\Sig}\g_I+\frac{1}{2\pi l_s}\left(\int_{S_a}\g_I\,\w\,A_{\Pi_a}-\int_{S^*_a}\g_I\,\w\,A_{\Pi_a^*}-\int_{S_b}\g_I\,\w\,A_{\Pi_b}+\int_{S^*_b}\g_I\,\w\,A_{\Pi_b^*}\right)\right ]
\ee
with $\Sigma$ a 3-chain such that $\p\Sig=\Pi_a-\Pi_b-\Pi_a^* +\Pi_b^*$.

\subsection{General case and $U(1)$ mixing from supergravity}

We can generalize these results to arbitrary configurations of parallel D7-branes with fluxes. Consider $K$ stacks of D7-branes wrapping divisors $S_\a$ and carrying magnetic fluxes $F_\a$ so we may associate a generalized submanifold $(S_a,\mathcal F_\a)$ to each of them. We find a massless $U(1)_X$ for every linear combination
\be
(S,\mathcal F)_X=\sum_{\a=1}^Kn_{X\a}(S_\a,\mathcal F_\a)
\ee
such that $(S,\mathcal F)^-_X\equiv\frac{1}{2}\left[(S,\mathcal F)_X-(S,\mathcal F)^*_X\right]$ is trivial in generalized homology. Thus, we may associate a chain $(\mathfrak S,\mathfrak F)_X$ such that
\be\label{ghom}
\hat \p(\mathfrak S,\mathfrak F)_X=(S,\mathcal F)^-_X.
\ee
Notice that, as our previous example shows explicitly, this chain may be the linear combination of different terms, namely
\be
(\mathfrak S,\mathfrak F)_X=\sum_i(\Sigma^{(i)},\mathcal F^{(i)})_X,
\ee
where the $\Sigma^{(i)}_X$ are not necessarily five-dimensional. Following what we did in the type IIA case, we fix the normalization by taking the following basis of open string $U(1)$'s
\be
\hat U(1)_\a=\frac{1}{L_\a}\sum_{\b=1}^Kn_{\a\b}U(1)_\b
\ee
with $n_{\a\b}$ and $L_\a$ as in (\ref{linX}).

In order to compute the kinetic mixing of a massless $\hat U(1)_X$ with the closed string sector we dimensionally reduce the action of a magnetic monopole of $\hat U(1)_X$ and look at the relevant couplings in 4d. In the simplest cases the monopoles are given by magnetized D5-branes wrapped on $W\times \Sigma$, with $\Sigma$ a 5-chain that connects the different D7-branes. However, in the general case these are given by wrapping bound states of different D$p$-branes, dubbed D-brane networks, on $W\times (\mathfrak S,\mathfrak F)_X$. By looking at the CS action of such network we find that the kinetic mixing with a closed string $U(1)_I$ is
\be\label{mibg}
f_{XI}=-\frac{i}{2l_s^4}\,j_{(\mathfrak S,\mathfrak F)_X}(\g_I),
\ee
where $\g_I$ is the harmonic 3-form that yields $U(1)_I$, and the associated current is given by
\be
j_{(\mathfrak S,\mathfrak F)_X}=\sum_i j_{(S^{(i)},\mathcal F^{(i)})_X}.
\ee

The current $j_{(\mathfrak S,\mathfrak F)_X}$ corresponds to the Lefschetz dual of the chain $(\mathfrak S,\mathfrak F)_X$ so the equation (\ref{ghom}) can be written in cohomology as
\be\label{zxc}
dj_{(\mathfrak S,\mathfrak F)_X}=j_{(S,\mathcal F)^-_X},
\ee
which is analogous to (\ref{backrG}). Since the current $j_{(S,\mathcal F)^-_X}$ has support only on the D7-branes we find that $j_{(\mathfrak S,\mathfrak F)_X}$ is an element of the generalized cohomology of $\mathcal M_6-S_X$, $H^{\bullet}(\mathcal M_6-S_X)$, with $S_X$ the union of the divisors wrapped by the D7-branes and their orientifold images. Following the discussion for the D6-branes we define the polyform $\varpi_X$ to be the representative of the class $[j_{(\mathfrak S,\mathfrak F)_X}]$ that is coclosed and integral, namely
\be\label{zxc2}
[\varpi_X]=[j_{(\mathfrak S,\mathfrak F)_X}]\in H^{\bullet}(\mathcal M_6-S_X), \quad d^*\varpi_X=0,\quad \int_{\mathcal M_6}\langle \varpi_X,\Upsilon\rangle\in\mathbb Z,
\ee
where $\langle \cdot,\cdot\rangle$ denotes the Mukai pairing and $\Upsilon$ is any polyform in $H^\bullet(\mathcal M_6,S_X,\IZ)$, meaning that $d\Upsilon=\iota^*_{S_X}\Upsilon=0$. The Mukai pairing is the natural generalization of the intersection pairing used in (\ref{condG}). Indeed, we may write eq.(\ref{condG}) as $\int\varpi\wedge \Upsilon\in\mathbb Z$ where $\Upsilon\in H^4(\mathcal M_6,\pi_{D6},\IZ)$ is the Lefschetz dual of $\Lambda_2\in H_2(\mathcal M_6-S_a\cup S_b,\IZ)$.

Using the polyform $\varpi_X$ we may write the mixing (\ref{mibg}) as an integral over $\mathcal M_6$
\be
f_{XI}=-\frac{i}{2l_s^4}\,j_{(\mathfrak S,\mathfrak F)_X}(\g_I)=-\frac{i}{2l_s^4}\,\int_{\mathcal M_6}\langle \gamma_I, j_{(\mathfrak S,\mathfrak F)_X}\rangle=-\frac{i}{2l_s^4}\,\int_{\mathcal M_6}\langle \gamma_I,\varpi_X\rangle
\label{genmix}
\ee
where we used the fact that $j_{(\mathfrak S,\mathfrak F)_X}$ and $\varpi_X$ only differ by an exact polyform which does not contribute to the integral since $\gamma_I$ is harmonic. Similarly to the case of D6-branes, we say that a linear combination of D-branes are linearly equivalent to zero if this combination is trivial in generalized homology and the corresponding massless $U(1)$ does not mix kinetically with the closed string sector.

\subsubsection*{An example}

Since these expressions are rather abstract, let us briefly illustrate it with the simple example analysed at the beginning of this section, namely two homotopic D7-branes $a$ and $b$ on $S_a\simeq S_b\simeq S$ with $[\mathcal F_a]=[\mathcal F_b]$ both in the bulk as well as in the cohomology of $S$ so we have that $\hat\p(\Gamma,\tilde{\mathcal F})=(S_a,\mathcal F_a)-(S_b,{\mathcal F_b})$. According to (\ref{zxc2}) we have that the polyform $\varpi$ associated to the massless $U(1)=\frac{1}{2}(U(1)_a-U(1)_b)$ satisfies
\be\label{tyu}
d\varpi_{(a-b)}=j_{(S_a,\mathcal F_a)}-j_{(S_b,\mathcal F_b)}
\ee
as well as the last two conditions therein. Because $j_{(S_\a,\mathcal F_\a)}$ contain the D7, D5 and D3-brane sources, they are polyforms and so is $\varpi_{(a-b)}$. More precisely we have that 
\be
\varpi_{(a-b)}\, =\, \varpi_1 + \varpi_3 + \varpi_5
\label{varpi7ab}
\ee
where 
\bea
d\varpi_1 & = & \delta(S_a) - \delta(S_b)\\
d\varpi_3 & = & \delta(S_a)\wedge\mathcal F_a-\delta(S_b)\wedge\mathcal F_b\\
d\varpi_5 & = & \delta(S_a)\wedge\oh\mathcal F_a^2-\delta(S_b)\wedge\oh\mathcal F_b^2
\eea
Following our general discussion in subsection \ref{ociia}, each of these $p$-forms $\varpi_i$ will have a decomposition of the form (\ref{Hodge1}). In particular we have that
\be
\varpi_1\, =\, \frac{1}{2\pi} d \left( {\rm ln} | h(S_a) |  -  {\rm ln} | h(S_b) |\right)
\label{varpi1}
\ee
where $h(S_\a)$ is the divisor function of $S_\a$. Recall that in a generic Calabi-Yau three-fold there will be no harmonic one-forms. Hence $\varpi_1$ will have no harmonic piece in the decomposition (\ref{Hodge1}) and linear equivalence will be trivially satisfied with respect to this piece of $\varpi$. This corresponds to the well known fact that two homologous divisors are always linearly equivalent in a Calabi-Yau manifold. The same applies to $\varpi_5$, and so the condition of linearly equivalence and the kinetic mixing  only depends on the piece $\varpi_3$.

Let us now assume that $\hat{\mathcal F}$ is a closed 2-form defined over the whole space $\mathcal M_6$ and such that $\hat{\mathcal F}|_{S_\a}=\mathcal F_\a$, $\a = a, b$.\footnote{Such 2-form may be obtained by trading the gauge fields on the D7-branes into the B-field as in \cite{dps14}.} We may then write (\ref{varpi7ab}) as
\be
\varpi_{(a-b)}\, =\, \varpi_1  \wedge e^{\hat{\mathcal F}}
\ee
with $\varpi_1$ as in (\ref{varpi1}). Applying now (\ref{genmix}) we obtain 
\be
f_{I(a-b)}=-\frac{i}{2l_s^6}\int_{\mathcal M_6}\gamma_I\w\varpi_3=-\frac{i}{2l_s^6}\int_{\mathcal M_6}\gamma_I\w\varpi_1\w\hat{\mathcal F}
\ee
which can be seen to be equivalent to our previous expression (\ref{kinbbo}).

\subsection{Application to F-theory GUTs}
\label{FGUTs}

While we have focused the discussion of this section in type IIB orientifold vacua, our main results can be extended to an F-theory setup along the lines of \cite{dps14}. In particular, they can be easily applied to an F-theory GUT model where the GUT gauge symmetry is broken by the presence of an hypercharge flux. In the present section we will do so, first obtaining the kinetic mixing of the hypercharge with bulk $U(1)$'s in a $SU(5)$ F-theory model. As this derivation is somewhat technical we present the result here
\be
f_{YI}=-\frac{5i}{2l_s^4}\,\left [ \int_\Sigma\gamma_I+\frac{1}{2\pi l_s}\int_{S_{GUT}}\gamma_I\wedge A_{GUT}\right ]
\label{mixtor}
\ee
where $\Sigma$ is a 3-chain such that $\p\Sigma=6\Pi$, with $[\Pi]$ Poincar\'e dual to the hypercharge flux class $[\frac{5}{6}F_Y]$, $dA_{GUT}=[\frac{5}{6}F_Y-\,\delta_{S_{GUT}}^2(\Pi)]$ and $\g_I$ is the 3-form that represents the RR $U(1)_I$ gauge symmetry.  As pointed out in the introduction, if a massless $U(1)_h$ of a hidden 7-brane sector mixes with $U(1)_I$ as well this will give rise to light particles with small irrational hypercharges. 

Nevertheless, even if the absence of such $U(1)_h$ a non-trivial mixing of the form could be relevant for F-theory GUT models, in the sense that the hypercharge normalization needed to absorb the kinetic mixing $\re f_{YI}$ will modify the relations between $\alpha_1$ and $\alpha_2$, $\alpha_3$ \cite{Redondo:2008zf}. This could be an interesting effect in view of the further corrections to the $SU(3) \times SU(2) \times U(1)_Y$ gauge coupling constants that arise from the 7-brane magnetization \cite{Donagi:2008ca,Blumenhagen:2008aw,Conlon:2009qa}. Another potential solution to this problem is to consider the case where the hypercharge has a mass mixing with a RR photon. As pointed out in \cite{cim11}, this may happen if the three-fold base $\CM_6$ contains torsional 2-cycles. We will apply our results to this case in the second part of this section.

\subsubsection*{Hypercharge mixing}

Consider 5 parallel 7-branes wrapping divisors $S_n$ in $\mathcal M_6$ with $n=1,\dots, 5$ and all of them homotopic to $S_{GUT}$ with vanishing B-field. Since the hypercharge generator within $SU(5)$ is given by\footnote{Notice that this normalization does not satisfy the conditions in the last section. However, it is the usual one for the hypercharge.} $Q_Y=\frac{1}{6}$diag$(2,2,2,-3,-3)$ we will  turn on fluxes $ F_n$ on all of them such that $ F_n=\frac{1}{3} F_Y$ for $n=1,2,3$ and $ F_n=-\frac{1}{2} F_Y$ for $n=4,5$ with $[ F_Y]$ trivial in the cohomology of $\mathcal M_6$ but non-trivial in $H^2(S_n)$
and such that $\frac{5}{6}[F_Y] \in H^2(S_n,\mathbb{Z})$ \cite{dw08}. Thus, when we bring the 7-branes back together this will translate into a flux $F_YQ_Y$ along the hypercharge generator that breaks $SU(5)\rightarrow SU(3)\times SU(2)\times U(1)_Y$ and keeps the hypercharge massless.

Given this data we may apply the results of the previous section to compute kinetic mixing with any closed string $U(1)_I$.
To each 7-brane we assign the generalized submanifold $(S_n,\mathcal F_n)$ and we have to look for the chain $(\mathfrak S,\mathfrak F)_Y$ that satisfies
\be\label{zxc}
\hat \p(\mathfrak S,\mathfrak F)_Y=2(S_1, F_1)+2(S_2,  F_2)+2(S_3, F_3)-3(S_4, F_4)-3(S_5, F_5).
\ee
According to eq.(3.23) we then have
\be
(S_n, F_n)=(S_n,0)+a_n(\Pi_n,0)+\hat \p\left [ -(\Gamma_n,\tilde{F}_n)+(\Gamma_n,\tilde{F}_{a_n\Pi_n}) \right ]
\ee
for all $n$. Here $\Pi_n\subset S_n$ is any 2-cycle such that $a_n[\Pi_n]=$P.D.$[F_n]$, with $a_n=\frac{2}{5}$ for $n=1,2,3$ and $a_n=\frac{3}{5}$ for $n=4,5$. See also eqs.(3.24-3.25). If we substitute this expression in (\ref{zxc}) we find that
\be
(\mathfrak S,\mathfrak F)_Y=(\Gamma,0)+(\Sigma,0)+\sum_{n=1}^5n_{Yn}\left [ -(\Gamma_n,\tilde{F}_n)+(\Gamma_n,\tilde{F}_{a_n\Pi_n})\right ],
\ee
modulo a generalized cycle. Here $n_{Yn}$ are the integers that define the hypercharge. Also, we have defined the 5-chain $\Gamma$ such that $\p\Gamma=\sum_n n_{Yn}S_n$ and the 3-chain $\Sigma$ that satisfies
 $\p\Sigma=\sum_n n_{Yn}a_n\Pi_n$. These exist given our initial hypotheses that $U(1)_Y$ is indeed massless.
 
The kinetic mixing with $U(1)_I$ is given by
\be
f_{YI}=-\frac{5i}{2l_s^4}\,j_{(\mathfrak S,\mathfrak F)_Y}(\g_I)
\ee
where the factor of 5 is due to the normalization of the hypercharge. The current $j$ is a sum of different contributions, namely
\be
j_{(\mathfrak S,\mathfrak F)_Y}=j_{(\Gamma,0)}+j_{(\Sigma,0)}+\sum_{n=1}^5n_{Yn}j_n.
\label{sumj}
\ee
where we have defined
\be
j_n=j_{(\Gamma_n,\tilde{ F}_{a_n\Pi_n})}-j_{(\Gamma_n,\tilde{ F}_n)}
\label{difj}
\ee
and which, using the results of appendix \ref{ap:Taub}, can be written as
\be
j_n(\gamma_I)=\frac{1}{2\pi l_s}\int_{S_n}\gamma_I\wedge A_{\Pi_n}
\ee
with $A_{\Pi_n}$ a 1-form on $S_n$ that satisfies
\be
d A_{\Pi_n}= F_n-\delta_{S_n}^2(a_n\Pi_n)
\ee
for $n=1,2,3$ we then have
\be
d A_{\Pi_n}=\frac{2}{5} \left [\frac{5}{6}F_Y-\delta_{S_n}^2(\Pi)\right ]
\ee
with $\Pi$ the Poincar\'e dual of $\frac{5}{6}F_Y$. For $n=4,5$ the corresponding equations reads
\be
d A_{\Pi_n}=-\frac{3}{5} \left [\frac{5}{6}F_Y -\delta_{S_n}^2(\Pi)\right ]
\ee
Putting everything together we find that
\be
f_{YI}=-\frac{5i}{2l_s^4}\,\left [ \int_\Sigma\gamma_I+\frac{1}{2\pi l_s}\sum_{n=1}^5 n_{Yn}\int_{S_n}\gamma_I\wedge A_{\Pi_n}\right ].
\ee
In order to get the mixing for the F-theory case we have to bring all the 7-branes together so $S_n=S_{GUT}$. This yields 
\be
f_{YI}=-\frac{5i}{2l_s^4}\,\left [ \int_\Sigma\gamma_I+\frac{1}{2\pi l_s}\int_{S_{GUT}}\gamma_I\wedge A_{GUT}\right ]
\ee
with $\p\Sigma=6\Pi$ and, using (\ref{sumj}) and (\ref{difj}), we find that $A_{GUT}=\sum_nn_{Yn}A_{\Pi_n}$ satisfies
\be
d A_{GUT}=6\left [F_Y- \frac{6}{5}\,\delta_{S_{GUT}}^2(\Pi)\right ].
\ee

As the RR $U(1)$s do not have any light charged states, one may simply absorb this mixing in a redefinition of the hypercharge coupling constant. Besides the implications for gauge coupling unification mentioned above, in a setup with low scale supersymmetry the gauginos of the hidden sector will be massive and may mix with the MSSM neutralinos which would lead to new signatures at the LHC, if SUSY is found, through different decay patterns. 

\subsubsection*{Implications for torsional hypercharge}

Following \cite{cim11}, let us now consider the case in which the hypercharge has a mass mixing with a bulk $U(1)_t$. For this to happen the 2-cycle class $[\Pi]$ Poincar\'e dual to the hypercharge flux $F_Y$ in $S_{GUT}$ must be a torsional 2-cycle of the three-fold base $\CM_6$. This means in particular that we have a non-trivial set of torsional cohomology classes in $\CM_6$, due to the identities 
\be
\text{Tor}H_2(\mathcal M_6,\mathbb Z)\simeq\text{Tor} H_3(\mathcal M_6,\mathbb Z)\simeq\text{Tor}H^3(\mathcal M_6,\mathbb Z)\simeq\text{Tor} H^4(\mathcal M_6,\mathbb Z)
\label{torclass}
\ee
Let us in particular assume that these groups are all equal to $\IZ_k$. Then we have the set of relations 
\be
d\omega^{\text{tor}} = k\,\b^{\text{tor}}\qquad \qquad d\a^{\text{tor}} = -k\,\tilde\omega^{\text{tor}}
\ee
where $\a^{\text{tor}}$, $\beta^{\text{tor}}$ are 3-forms of $\CM_6$ which are also eigenforms of the Laplacian, and we have the normalization 
\be
\int_{\mathcal M_6} \a^{\text{tor}}\w\, \b^{\text{tor}}\, =\,  \int_{\mathcal M_6} \omega^{\text{tor}}\w \,\tilde\omega^{\text{tor}}=1
\ee
We then expand the RR potential $C_4$ on these non-harmonic forms and obtain
\be
C_4= A_1^t\wedge \re\, \g^{\text{tor}}-V_1^t\wedge \im \, \g^{\text{tor}} +  \re f_{tt}\,C_{2,t}\wedge \omega_i^{\text{tor}} -\textrm{Re}(T^t)\tilde{\omega}^{\text{tor}}
\ee
where just like for harmonic forms, we consider the following combination
\be
\gamma^{\text{tor}}\, =\, \a^{\text{tor}} + i f_{tt} \b^{\text{tor}} 
\ee
with $f_{tt}$ chosen so that $\g^{\text{tor}}$ is a (2,1)-form. The 3-form $\gamma^{\text{tor}}$ represents the bulk $U(1)_t$ that corresponds to the torsion group (\ref{torclass}), and $f_{tt}$ its gauge kinetic function. Following \cite{cim11}, it is easy to see that the kinetic mixing between the hypercharge and this $U(1)_t$ will also be given by (\ref{mixtor}) with basically $\g_I$ replaced by $\g^{\rm tor}$. More precisely, since $[\Pi]$ is non-trivial in $H_2(\CM_6,\IZ)$ but $k[\Pi]$ is, we can consider a monopole with charge $k$ under the 7-brane $U(1)$, and by analyzing its worldvolume theory we get the expression
\be
\re f_{Yt}=\frac{5}{2l_s^4}\, \frac{1}{k} \left [ \int_\Sigma \im \gamma^{\text{tor}} +\frac{1}{2\pi l_s}\int_{S_{GUT}}\hspace*{-.4cm} \im \gamma^{\text{tor}} \wedge A_{GUT}\right ]\, =\, \frac{5}{2l_s^4} \re f_{tt} \int_{S_{GUT}} \hspace*{-.4cm} \omega^{\rm tor} \wedge F_Y
\ee
where $\Sigma$ is a 3-chain ending on $k$ copies of $\Pi$, and $dA_{GUT}=[k\frac{5}{6}F_Y-\, \sum_{i=1}^k \delta_{S_{GUT}}^2(\Pi_i)]$. We could have also obtained this expression by direct dimensional reduction of the Chern-Simons action $\int C_4 \wedge F \wedge F$ of the GUT 7-brane, which gives
\be
\int_{\IR^{1,3}} F_2^Y \wedge (dV_1^t + k C_{2,t}) \left[ \frac{5}{2l_s^4}\re f_{tt} \int_{S_{GUT}} \hspace*{-.4cm} \omega^{\rm tor} \wedge F_Y \right]
\ee

Since the hypercharge flux induces a torsion class $[\Pi]$ in $\CM_6$ we will have a relation of the form P.D.$[\Pi] = k_Y [\tilde \om^{\text{tor}}]$, where Poincar\'e duality is performed in $\CM_6$. Then, by the results of \cite{cim11} we have that the hypercharge and the bulk $U(1)_t$ have a mass mixing, and that the massless $U(1)_{\tilde Y}$ generator is given by the linear combination 
\be
A_{1}\, =\, \cos \theta \tilde A_Y -\sin \theta \tilde A_t\,, \quad \sin \theta = \frac{k_Y g_Y}{\sqrt{k_Y^2g_Y^2 + k^2 g_t^2}}\,,
\ee
in terms of the gauge bosons $\tilde A_Y = A_Y g_Y^{-1}$ and $\tilde A_t = A_t g_t^{-1}$ with canonical kinetic term.
The Lagrangian in 4d contains the terms
\be
\mathcal L\supset -\frac{2\pi}{l_s^2}\big [ \re f_{YY}\,F_Y\w*F_Y +\re f_{tt}\,F_t\w*F_t+ 2\re f_{Yt}\,F_Y\w*F_t \big ]
\ee
so we may readily compute the kinetic function of the massless eigenstate by rotating to the mass eigenstate basis, namely
\be
\mathcal L\supset -\frac{2\pi}{l_s^2}\re f_{11}\,F_{1}\w* F_{1},
\ee
with
\be
\re f_{11}= \,\cos^2 \theta \,\re f_{YY}- 2 \cos \theta \sin \theta  \,\re f_{Yt} + \sin^2 \theta\, \re f_{tt}.
\ee
With this normalization the field $A_{1}$ couples to the charges states with a coupling constant $\cos \theta g_Y$ so we should redefine $A_{\tilde Y}=\cos \theta A_1$ which has a gauge kinetic function
\be\begin{split}
\re f_{\tilde Y\tilde Y}&= \,\re f_{YY}- 2  \tan  \theta \,\re f_{Yt} + \tan^2\theta \, \re f_{tt}\\
&=\frac{5}{3 \alpha_G}- 2 \frac{k_Y}{k} \,\re f_{Yt} + \frac{k_Y^2}{k^2 \alpha_t^2}\,.
\end{split}\ee
This generalizes the result obtained in section 5 of \cite{cim11}. As pointed out there, these corrections may explain the small discrepancy in gauge coupling unification found in standard F-theory GUTs. Our generalization including the kinetic mixing $f_{Yt}$ provides an even more flexible scheme to fix such discrepancy.

\section{Conclusions}
\label{conclu}

Motivated by their implications for the existence of milli-charged particles, in this paper we have analyzed the kinetic mixing between closed string $U(1)$'s and those associated with D-branes in Type II orientifolds. Even if there are no light states charged under closed string $U(1)$'s, they may induce the presence elf milli-charged particles if they have a kinetic mixing with open string $U(1)$'s in a visible and hidden sectors simultaneously. 

We have started by computing the open-closed mixing in type IIA models of intersecting D6-brane models by simple dimensional reduction of the Chern-Simons (CS) action of two parallel or homotopic D6-branes. We have seen how this mixing can be rephrased in terms of the Witten effect for the open string magnetic monopoles. Such vantage point not only allows to derive the mixing in more involved configurations, but also shows the importance of the relative (co)homology groups in this kind of constructions. For instance, in type IIA compactifications, the magnetic monopoles of massless $U(1)$'s are D4-branes wrapped on the 4-chain that connects the relevant D6-branes. Therefore, they are classified by the fourth homology group of the compactification space relative to the position of the D6-branes, which already includes the usual homology that classifies the closed string monopoles. By performing the reduction of the CS action of such monopoles we have computed their electric charge under the bulk $U(1)$'s which is precisely related to the kinetic mixing. While this is done purely in a perturbative regime, it matches the result obtained from the lift to M-theory \cite{cim11}.

Moreover, we have seen that the mixing has a nice geometrical interpretation in terms of linear equivalence of submanifolds \cite{Hitchin99}. This is a criterion used to compare a linear combination of cycles by associating them a bundle or, more generally, a gerbe. When this combination is homologically trivial, the corresponding gerbe can be chosen flat and it is fully characterized by its holonomy. When such holonomy is trivial, the $U(1)$ associated to the cycles does not mix kinetically with the bulk sector, as we have shown. In fact, this criterion captures the mixing due to geometrical deformations moduli of the D6-branes but not the effect of their Wilson line moduli. One can nevertheless extend the concept of linear equivalence to submanifolds that carry gauge bundles on them, dubbed D-brane linear equivalence. 

Since the mixing can be obtained from measuring the charges of a $U(1)$ monopole under other $U(1)$'s, it can be computed by dimensionally reducing the CS action of such monopole. This leads us to an expression for the mixing which is an integral over the $p$-chain wrapped by the monopole in the internal dimensions. Using Lefschetz duality, one can nevertheless write it as an integral over the whole compactification space, yielding a formula that closely resembles the mixing in the closed string sector.
 
Following the same strategy, we have analyzed the same problem in the context of magnetized D7-brane compactifications. This case is particularly interesting since it is closely related to F-theory GUT models. The major difference with type IIA is the presence of worldvolume fluxes. In particular, these contribute to the St\"uckelberg couplings so it is not enough to have homologous cycles to get a massless $U(1)$. Nevertheless, when expressed in terms of generalized geometry, the stategy pursued in the previous case still applies. Indeed, we find a massless $U(1)$ for every linear combination of magnetized D7-branes that is trivial in generalized homology and its magnetic monopole is given by wrapping a bound state of D-branes (a D-brane network in the language of  \cite{em07}) in the corresponding generalized chain. Again, using the Witten effect one can compute the kinetic mixing with the closed string sector, and understand this mixing in terms of D-brane linear equivalence.

We have applied our results to the case of F-theory GUTs when the unification group is broken by a hypercharge flux. This case is particularly subtle since in order to have a massless $U(1)_Y$ the hypercharge flux must be trivial in the cohomology of the bulk and non-trivial in the worldvolume of the 7-brane. This subtlety is relevant when identifying magnetic monopoles, which arise as an element of a generalized relative homology group.\footnote{In this sense, our findings refine the monopole construction in \cite{dw08}.} Once the monopoles are identified one may compute the $U(1)_Y$ mixing with bulk $U(1)$'s, which may lead to phenomenologically interesting consequences besides the presence of milli-charged particles. Indeed, either a kinetic or mass mixing induces a redefinition of the hypercharge coupling that could be relevant for gauge coupling unification. Finally, in an MSSM scheme, it may induce a mixing between hidden gauginos, (typically massive upon SUSY breaking) and the MSSM neutralinos, and this could lead to new signatures at the LHC. 

The fact that the mixing between $U(1)$'s can be written as an integral over the whole compactification space upon considering relative (co)homology groups instead of the usual cohomology is very suggestive, since then the expressions closed-closed and open-closed mixing look alike. This is perhaps not too surprising, since in the M/F-theory uplift of type II compactifications all the $U(1)$'s are classified by the same topological group. It would be then interesting to examine in more detail the role of relative (co)homology in type II compactifications with D-branes. Furthermore, one can try to generalize this method to more general setups in several directions. For instance, the language of generalized complex geometry is well suited to compactifications with non-vanishing background fluxes and their mirror symmetry duals. Finally, one may wonder if the use of relative cohomology groups may lead to an alternative way to compute kinetic mixing between open string $U(1)$'s, a problem which we plan to address in the near future.

\bigskip

\bigskip

\centerline{\bf \large Acknowledgments}

\bigskip

We would like to thank Pablo~G.~C\'amara, I\~naki~Garc\'ia-Etxebarria, Luis~E.~Ib\'a\~nez, V\'ictor Mart\'in-Lozano, Pablo Soler and Angel~M.~Uranga for useful discussions. 
This work has been partially supported by the grant FPA2012-32828 from the MINECO, the REA grant agreement PCIG10-GA-2011-304023 from the People Programme of FP7 (Marie Curie Action), the ERC Advanced Grant SPLE under contract ERC-2012-ADG-20120216-320421 and the grant SEV-2012-0249 of the ``Centro de Excelencia Severo Ochoa" Programme. F.M. is supported by the Ram\'on y Cajal programme through the grant RYC-2009-05096. D.R. is supported through the FPU grant AP2010-5687. G.Z. is supported through a grant from ``Campus Excelencia Internacional UAM+CSIC".

\clearpage

\appendix


\section{Linear equivalence of $p$-cycles}
\label{Leq}

In this appendix we will review the definition of linear equivalence as given in \cite{Hitchin99} for general submanifolds. We will start by reviewing the concept of $p$-gerbe with connection which will enter directly in the definition of linear equivalence and then give the definition of linear equivalence of
submanifolds.
\subsection{From bundles to $p$-gerbes}

Roughly speaking a $p$-gerbe is a generalization in higher dimension of a line bundle. In order to motivate its definition we start by giving three equivalent characterizations of a line bundle $\mathcal{L}$ on a manifold $X$:
\begin{itemize}
\item[-] A cohomology class in $H^2(X,\mathbb{Z})$,
\item[-] A real codimension 2 submanifold $M$ of $X$,
\item[-] An element in the \v Cech cohomology group $\check{H}^1(X,U(1))$.
\end{itemize}

The cohomology class characterizing $\mathcal{L}$ is its first Chern class $c_1(\mathcal{L})$ while the real codimension 2 submanifold is the Poincar\'e dual of $c_1(\mathcal{L})$. Finally the element in $\check{H}^1(X,U(1))$ specifies the transition functions of the bundle, namely taking an open cover $\{ \mathcal{U}_
\alpha\}$ of $X$ such that $\mathcal{L}$ is trivial over each set $\mathcal{U}_\alpha$ given $g \in \check{H}^1(X,U(1))$ we get the functions
\be
g_{\alpha \beta} :\ \mathcal{U}_\alpha \cap \mathcal{U}_\beta \rightarrow U(1)\,,
\ee
such that $g_{\alpha \beta}(x)  g_{\beta \alpha}(x)=1$ in $ \mathcal{U}_\alpha \cap \mathcal{U}_\beta$ and furthermore
\be
g_{\alpha \beta}(x) g_{\beta \gamma}(x) g_{\gamma\alpha }(x) =1\,, \quad \forall x \in \mathcal{U}_\alpha \cap \mathcal{U}_\beta \cap \mathcal{U}_\gamma\,.
\ee

The natural generalization of the previous characterizations of a line bundle are the following ones:

\begin{itemize}
\item[-] A cohomology class in $H^{p+2}(X,\mathbb{Z})$,
\item[-] A real codimension $p+2$ submanifold $M$ of $X$,
\item[-] An element in the \v Cech cohomology group $\check{H}^{p+1}(X,U(1))$.
\end{itemize}

We will take one of the three equivalent characterizations as a definition of a $p$-gerbe. We now will endow $p$-gerbes with a connection in a way similar to how we endow line bundles with a connection and relate the curvature of this connection to the cohomology class in $H^{p+2}(X,\mathbb{Z})$.

\subsection{Connections on $p$-gerbes}

It is again useful to start recalling how a connection is built on a line bundle. Given a open cover $\{\mathcal{U}_\alpha\}$ of $X$ a connection a connection on a line bundle $\mathcal{L}$ with transitions functions $g \in \check{H}^1(X,U(1))$ 
is a set of 1-forms $A_{\alpha}$ defined on $\mathcal{U}_{\alpha}$ that on double intersections $\mathcal{U}_{\alpha} \cap \mathcal{U}_\beta$ satisfy
\be
i( A_{\beta}-A_{\alpha})= g_{\alpha \beta}^{-1} d g_{\alpha \beta}\,.
\ee
In particular since $d( g_{\alpha \beta}^{-1} d g_{\alpha \beta})=0$ we have that there is a global closed 2-form (the curvature of the bundle) satisfying
\be
F|_{\mathcal{U}_\alpha} = d A_{\alpha}\,.
\ee
We can now adapt this procedure to the case of a $p$-gerbe. Let us call $C^q(\mathcal{U},\mathcal{F})$ the set of $q$ \v Cech cochains with values in a sheaf $\mathcal{F}$ for the open covering $\mathcal{U}$. Then given the transition functions of the $p$-gerbe $\mathcal{G}$ 
we can build the element $\varpi^{(1)} \in C^{p}(\mathcal{U},\Omega^1(X))$ satisfying
\be
(\delta \varpi^{(1)})_{\alpha_1 \dots \alpha_p} =( g^{-1} d g)_{\alpha_1 \dots \alpha_p}\,,
\ee
where $g \in \check H ^{p+1}(X,U(1))$ are the transition functions of the gerbe $\mathcal{G}$. We can iteratively arrive at the definition of the curvature of the connection
\be
(d \varpi^{(q)})_{\alpha_1 \dots \alpha_{p-q}} = (\delta \varpi^{(q+1)})_{\alpha_1 \dots \alpha_{p-q}}\,,
\ee
where at each stage we have $\varpi^{(q)} \in C^{p-q+1}(\mathcal{U},\Omega^q(X))$. This can be repeated until we arrive at $\varpi^{(p+1)}$ which is an element of $C^0(\mathcal{U},\Omega^{p+1}(X))$, we define $G/2\pi= d\varpi^{(p+1)}$ to be the curvature of the $p$-gerbe. It is a 
globally defined and closed $p+2$ 
form 
whose
class $[G]/2\pi \in H^{p+2}(X,\mathbb{R})$ is the image of the characteristic class of the gerbe $\mathcal{G}$ under the natural inclusion $i : H^{p+2}(X,\mathbb{Z}) \rightarrow H^{p+2}(X,\mathbb{R})$.

We can give a direct construction of the connection on a $p$-gerbe associated to a codimension $p+2$ submanifold and this will be very important in the definition of linear equivalence of submanifolds. Given a codimension
 $p+2$ submanifold $M$ we will denote 
its Poincar\'e dual $p+2$-form as $\delta(M)$. We now would like to find a $p+1$ $\varpi$ form that will be the connection on our $p$-gerbe, this will be a 0 \v Cech cochain that is defined by the following differential equations
\be
d \varpi ^{(p+1)}_{\alpha} = \delta(M)|_{\mathcal{U_{\alpha}}} \,.
\ee

This does not specify completely the connection for the addition of a closed form to it does not alter the previous differential equation. We can partially fix this ambiguity asking for the two conditions
\be\label{eq:conn}
d^* \varpi ^{(p+1)} = 0\,, \quad \int_{\Lambda_{p+1}} \varpi ^{(p+1)} \in \mathbb{Z}\,,
\ee
where $\Lambda_{p+1} \in H_{p+1}(X\setminus M,\mathbb{Z})$. These conditions will still be satisfied if we add to $\varpi $ an integral harmonic form but this ambiguity is not important in characterizing the connection on the $p$-gerbe. Once these conditions are imposed we see that the connection has the 
following Hodge decomposition
\be
\varpi ^{(p+1)} = \omega  + d^* H\,,
\ee
where $\omega $ the harmonic part of the connection and $d d^* H =  \delta (M)$. We next characterize the data specifying the connection.
As we already have the notion of curvature $G/2\pi \in H^{p+2}(X,\mathbb{Z})$ of a $p$-gerbe we only need to adapt the definition of holonomy which will take values in $H^{p+1}(X,\mathbb{R})
/H^{p+1}(X,\mathbb{Z})$. We characterize the holonomy of a $p$-gerbe as follows: take a non trivial $p+1$ cycle $\Sigma_{p+1}$ and the define the holonomy of the connection around it as
\be
hol(\varpi ^{(p+1)},\Sigma_{p+1}) = \mathrm{exp}\left(2 \pi i \int_{\Sigma_{p+1}} \omega \right)\,,
\ee
Choosing a basis of non trivial $p+1$ cycles we get a complete characterization of the holonomy of the connection. Since the holonomy and curvature data completely specify the connection
we have that a connection on a gerbe is trivial if both its holonomy and its curvature are zero.

\subsection{Linear equivalence of submanifolds}

We now give the definition of linear equivalence of submanifolds. Appearance of gerbes in the definition of linear equivalence is quite natural for every gerbe we have an associated submanifold. We say that two submanifolds 
$M$ and $N$ are linearly equivalent if the gerbe $\mathcal{G}_M \mathcal{G}_N^{-1}$ has a trivial connection.\footnote{Given a gerbe $\mathcal{G}$ we define its dual $\mathcal{G}^{-1}$ to be the gerbe whose transition functions are the inverse of the ones of the gerbe $\mathcal{G}$ Moreover we can introduce
a product in the space of $p$-gerbes defining the product of two $p$-gerbes $\mathcal{M}$ and $\mathcal{N}$ with transitions functions $g_\mathcal{M}$ and $g_\mathcal{N}$ to be the $p$-gerbe with transition functions $g_\mathcal{M} g_\mathcal{N}$. We shall call the product gerbe simply $\mathcal{M}\mathcal{N}
$. Note that the notion of dual and product agree with the known ones in the case of line bundles.}. We can easily extend this definition to linear combinations of submanifolds as it is usually done in the case of divisors: two linear combinations of submanifolds $M= \sum_i a_i M_i$ and $N= \sum_i b_i N_i$ are
 linearly equivalent if the connection on the gerbe $\prod_{i} \mathcal{G}^{a_i}_{M_i}\prod_{j} \mathcal{G}^{-b_j}_{N_j}$ has a trivial connection.
 
We now would like to characterize when two submanifolds of codimension $p+2$ are linearly equivalent, the more general case of a linear combination can be similarly understood.
The $p$-gerbe  $\mathcal{G}_M \mathcal{G}_N^{-1}$ has a connection satisfying the following differential equation
\be\label{eq:dA}
d \varpi _{\alpha} = 2\pi \left[\delta(M)-\delta(N)\right]|_{\mathcal{U_{\alpha}}} \,.
\ee
A first condition that we need to impose on the connection on the gerbe in order to be trivial is that its curvature is zero, this happens when the two submanifolds are in the same homology class so that the right hand side of (\ref{eq:dA}) is an exact form and the connection $\varpi$ is globally well defined.
All we need to check is that the bundle has trivial holonomy and this happens if the harmonic part of the connection $\varpi$ is form with integer periods.
If we impose the conditions (\ref{eq:conn}) we see that linear equivalence amounts to the following equivalent conditions on the connection on the gerbe $\mathcal{G}_M \mathcal{G}_{N}^{-1}$
\be
\omega \in H^{p+1}(X,\mathbb{Z})\quad  \Leftrightarrow \quad d^*H \in H^{p+1}_c(U,\mathbb{Z})
\ee
where $U \equiv X \setminus (M \cup N).$\footnote{We need to remove $M$ and $N$ from $X$ because the form $d^*H$ has poles on these submanifolds.}
 We can moreover give following \cite{Hitchin99} a further characterization of linear equivalence of submanifolds. We focus again on the case of two submanifolds but this can be easily generalized to a linear combination of submanifolds. We will show the following: two 
submanifolds $M$ and $N$ of codimension $p+2$ are linearly equivalent if and only if their homology classes agree and moreover
\be
\int_{\Gamma} \theta \in \mathbb{Z}\,,
\ee
where $\p \Gamma = M-N$ and $\theta$ is an harmonic form with integral cohomology class. This is quite easy to show: let us start by integrating a general form with compact support on $\Gamma$
\be
\int_\Gamma \alpha = \int_U \alpha \wedge \gamma
\ee
where we called $\gamma$ the Poincar\'e dual of $\Gamma$, we write its Hodge decomposition as $\gamma = A + dB + d^*C$.
 We can see that the coclosed part of $\gamma$ agrees with the coexact part of our connection on the gerbe $\mathcal{G}_M \mathcal{G}_{N}^{-1}$, in fact if we integrate an exact form on $\Gamma$ we get
\be
\int_{\Gamma} d \beta = \int_{\p \Gamma} \beta = \int_{M} \beta - \int_N \beta\,,
\ee
but we also have that
\be
\int_{\Gamma} d \beta = \int_U d \beta \wedge \gamma =  \int_U \beta \wedge d \gamma = \int_U \beta \wedge d d^*C\,, 
\ee
so that in the end we get that $d d^*C = \delta (M) -\delta (N) = d d^* H$. Moreover we have that $\Gamma$ is an integral cycle and if we integrate a closed form with compact support and integral cohomology class on it we get
\be
\int_\Gamma \phi = \int_U \phi \wedge (A + dB +  d^* C) =  \int_U \phi \wedge (A  +  d^* H) \in \mathbb{Z}\,,
\ee
that implies that $d^*H$ is an integral form if and only if $\int_U \phi \wedge A$ is integral. Since the compactly supported closed form with integral class $\phi$ is cohomologous to an integral harmonic form $\theta$ defined on the whole manifold $X$ we have that
\be
\int_U \phi \wedge A = \int_X \theta \wedge A = \int_\Gamma \theta\,,
\ee
which implies that
\be
\int_U \phi \wedge A \in \mathbb Z \, \iff \int_\Gamma \theta \in \mathbb Z\,.
\ee

\section{M-theory lift of D6-branes}
\label{ap:Taub}

In this appendix we show how the harmonic 2-forms $\varpi\in H^2(\mathcal M_6-\{\pi_{D6}\},\mathbb Z)$ appear from the M-theory lift of a configuration of D6-branes. Here $\pi_{D6}$ is the set of all branes.

The M-theory lift of $N$ parallel D6-branes in flat spacetime is given by a purely geometric configuration. More explicitly, we have that the lift corresponds to the space $\mathbb R^{1,3}\times \mathbb R^3\times {\bf TN}_N$ where ${\bf TN}_N$ is the N-center Taub-NUT space with metric 
\be
ds^2=Vd\vec r \cdot d\vec r+\frac{1}{V}(d\psi+\vec\omega\cdot d\vec r)^2
\ee
with $\vec r\in \mathbb R^3$ and $\psi\sim \psi+ 4\pi$. Also,
\be
V=\frac{1}{R}+\sum_{\a=1}^N V_\a,\qquad V_\a=\frac{1}{|\vec r-\vec a_\a|}
\ee
and
\be
\vec \omega = \sum_{\a=1}^N\vec \omega_\a,\qquad \vec \nabla V_\a=\vec \nabla\times \vec \omega_\a.
\ee
This geometry is a ${\bf S}^1$ fibration over $\mathbb R^3$ and the radius of the fiber is given by the $g_{\psi\psi}$ component of the metric, namely, $\frac{1}{V}$. Near the location of the D6-branes $\vec a_\a$ the radius goes like $\frac{1}{V_\a}=|\vec r-\vec a_\a|$ so the fiber shrinks. On the other hand, asymptotically far away from them it has constant radius $R$ which determines the string coupling constant in Type IIA, namely $g_s^{IIA}=\frac{R}{\sqrt {\a'}}$. 

An alternative way to understand this space is to regard it as the total space of a $U(1)$ gauge bundle over $\mathbb R^3$ with $N$ magnetic monopoles at $\vec a_\a$. Then, the one-forms $\omega_\a\equiv\vec \omega_\a\cdot d\vec r$ are the gauge fields associated to the monopole $\a$. Thus, their field strengths are given by $F_\a=d\omega_\a$ and satisfy the equations,
\be\label{bianchi}
dF_\a=-4\pi\delta_3(\vec r-\vec a_\a),\qquad d*F=0.
\ee

There are $N$ cohomologically independent harmonic normalizable 2-forms that, when reducing the $C_3$ potential, yield the gauge fields associated with the D6-branes. These are \cite{ruback86}
\be
4\pi\Omega_\a=d\chi_\a,\qquad \chi_\a=\frac{V_\a}{V}(d\psi+\omega)-\omega_\a.
\ee
which are orthogonal, i.e. $\int\Omega_\a\,\w\,\Omega_\b=R^2\delta_{\a\b}$. It is useful to take the linear combinations
\be
\Omega_{\a\b}=\Omega_\a-\Omega_\b,\qquad \Omega_{CM}=\sum_{\a=1}^N\Omega_\a
\ee
since $\Omega_{\a\b}$ has compact support while the center of mass 2-form $\Omega_{CM}$ does not.

For every harmonic 2-form $\Omega_{\a\b}$ with compact support there is a non-trivial 2-cycle in homology $\pi_{\a\b}$ (which is compact by definition). Topologically, these cycles are 2-spheres given by $\pi_{\a\b}\simeq\pi^{-1}(I_{\a\b})$ where $\pi: {\bf TN}_N\rightarrow \mathbb R^3$ is the projection of the fibration and $I_{\a\b}$ is a path in the base that connects $\vec a_{\a}$ and $\vec a_{\b}$. When we take the limit $R\rightarrow 0$ these 2-cycles become 1-chains in the base which are just the paths $I_{\a\b}$. In Type IIA the M-theory circle disappears from the description so these can be regarded as 1-cycles in $H_1(\mathbb R^3,P)$ where $P$ is the set of points $\vec a_{\a}$. Thus, by Poincar\'e-Lefschetz duality there should be $N-1$ independent 2-forms in $H^2(\mathbb R^3-P)$. Let us see this explicitly.

Consider the 2-forms $\Omega_{\a\b}$ in ${\bf TN}_N$, namely
\bea
4\pi\Omega_{\a\b}&=&d\left ( \frac{V_\a-V_\b}{V}(d\psi+\omega) \right )-d(\omega_\a-\omega_\b)\\\nonumber
&=&-\frac{R^2(V_\a-V_\b)}{R^2V^2}dV\w(d\psi+\omega)+\frac{R}{RV}d((V_\a-V_{\b})(d\psi+\omega))-d(\omega_\a-\omega_\b).
\eea
In the limit $R\rightarrow 0$ we have $RV\rightarrow 1$, so only the last term survives which yields
\be
\Omega_{\a\b}\,\,\overset{R\rightarrow 0}{\longrightarrow}\,\,\tilde \Omega_{\a\b}=\frac{1}{4\pi}d(\omega_\b-\omega_\a).
\ee
Using eq.(\ref{bianchi}) we see that $d\tilde\Omega_{\a\b}=\delta_3(\vec r-\vec a_\a)-\delta_3(\vec r-\vec a_\b)$ so it is a closed 2-form in $\mathbb R^3-P$ and $d*\tilde\Omega_{\a\b}=0$. One could think that since $\tilde\Omega_{\a\b}$ is given by the exterior derivative of $\omega_\b-\omega_\a$ it is trivial in cohomology, however, the one-forms $\omega_\a$ are not globally well-defined, instead they have to be defined in different patches and glued together using a gauge transformation. Thus, an alternative way of writing this 2-form is $\tilde\Omega_{\a\b}=d^*H_{\a\b}$ where $H_{\a\b}$ is a globally well-defined 3-form with a singularity of the form $\frac{1}{r}$ at the points $\vec a_\a$ and $\vec a_\b$. Thus, putting everything together we have that
\be
\tilde\Omega_{\a\b}=d^*H_{\a\b},\qquad d\tilde \Omega_{\a\b}=\delta_3(\vec r-\vec a_\a)-\delta_3(\vec r-\vec a_\b),\qquad d^*\tilde\Omega_{\a\b}=0.
\ee
This shows that $\tilde \Omega_{\a\b}$ are precisely the 2-forms $\varpi^{(\a-\b)}$ in the main text. Since this is a non-compact toy model there is no harmonic 2-form in $\tilde\Omega_{\a\b}$ which means that in flat space any two points are linearly equivalent. When embedded in a compact model such harmonic 2-form must be added to $\tilde \Omega_{\a\b}$ to ensure that it is integral.


\section{Generalized homology}
\label{gh}

In this appendix we give the basic definitions of the generalized homology introduced in \cite{em07}. We refer the reader to \cite{em07} for a detailed discussion on the subject.

The RR charges of D-branes in the presence of a NSNS $H$-flux are classes in twisted K-theory but it is generally very difficult to compute them in concrete examples. On the other hand, generalized homology captures some aspects of the RR charges beyond usual homology and is much easier to work with.

In the context of generalized complex geometry a Dp-brane can be described as a submanifold $\Sigma$ carrying a gauge bundle $\mathcal F=F+B$ which is a generalized submanifold $(\Sigma, \mathcal F)$.\footnote{We restrict ourselves to abelian D-branes.} In other to define a homology theory for these objects we need to define chains and a boundary operator that squares to zero.

Consider a Dp-brane wrapping a $(p+1)$-submanifold with a $U(1)$ gauge field that may have a Dirac monopole on a $(p-2)$-submanifold $\Pi\subset \Sigma$ which must satisfy $\p\Pi\subset\p\Sigma$. The field strength $\mathcal F_{\Pi}$ thus satisfies
\be
d\mathcal F_{\Pi}=H|_\Sigma+\delta_{\Sigma}(\Pi)
\ee
in the presence of a $H$-flux. Since $\mathcal F_{\Pi}$ should be globally well-defined we have that
\be
\text{P.D.}[H|_\Sigma]+[\Pi]=0.
\ee
Thus, the pair $(\Sigma, \mathcal F_{\Pi})$ is a generalized submanifold and a generalized chain is defined to be a formal sum of these pairs. We restrict the sum to contain only even or odd dimensional submanifolds as suggested by Type IIB and IIA string theories respectively.

We define the generalized boundary operator in such a way that that its action is dual to the action of the $H$-twisted exterior derivative on forms. Therefore, inspired in the CS action for a D-brane, we associate a current $j_{(\Sigma,\mathcal F_{\Pi})}$ to each chain by
\be
j_{(\Sigma,\mathcal F_{\Pi})}(C)\equiv\int_\Sigma C|_\Sigma\,\w\,e^{\mathcal F_\Pi}
\ee
where $C$ is an arbitrary polyform of definite parity. The derivative $d_H=d+H\,\w\,$ acts on the current in the following way
\be
(d_Hj_{(\Sigma,\mathcal F_{\Pi})})(C)=\int_\Sigma d_HC|_\Sigma\,\w\,e^{\mathcal F_\Pi}=\int_{\p\Sigma} C|_\Sigma\,\w\,e^{\mathcal F_\Pi|_{\p\Sigma}}-\int_\Pi C|_\Pi\,\w\,e^{\mathcal F_\Pi|_\Pi}
\ee
where we used Stokes' theorem. Thus,
\be
d_Hj_{(\Sigma,\mathcal F_{\Pi})}=j_{(\p\Sigma,\mathcal F_\Pi|_{\p\Sigma})}-j_{(\Pi, \mathcal  F_\Pi|_\Pi)}
\ee
so we define the generalized boundary operator $\hat\p$ by imposing
\be
d_Hj_{(\Sigma,\mathcal F_{\Pi})}=j_{\hat \p(\Sigma,\mathcal F_{\Pi})}
\ee
which leads to
\be
\hat\p(\Sigma,\mathcal F_\Pi)\equiv(\p\Sigma,\mathcal F_\Pi|_{\p\Sigma})-(\Pi, \mathcal F_\Pi|_\Pi).
\ee
One can check that $\hat \p^2=0$ which allows to define the generalized homology as Ker$(\hat\p)/$Im$(\hat\p)$.

In order to preserve RR gauge invariance D-branes can only wrap generalized chains that are closed \cite{em07}. Then for any two generalized cycles that are in the same homology class there is a physical process that connects them although it may not be energetically favorable. In order to define the energy of a generalized cycle one can introduce a generalized calibration \cite{em07}. Of particular relevance for our discussion is the effect of dissolving D(p-2)-branes in Dp-branes which is nicely captured in this formalism since both situations correspond to different representatives of the same homology class as is shown in the main text.


\section{Details on the computation of $j_{(\mathfrak S,\mathfrak F)}$}
\label{cal}

Here we derive the equation (\ref{jfrak}) starting with the generalized chain $(\mathfrak S,\mathfrak F)$ and show it has all the desired properties.

First, one can check that a change in $S'_a$, $\F'_a$ or $B_a$ corresponds to choosing a different generalized chain $(\mathfrak S',\mathfrak F')=(\mathfrak S,\mathfrak F)+\hat\p(\mathfrak s,\mathfrak f)$ so the associated current is $j_{(\mathfrak S',\mathfrak F')}=j_{(\mathfrak S,\mathfrak F)}+d\,j_{(\mathfrak s,\mathfrak f)}$. Since $\g_I$ is harmonic we have that $d\,j_{(\mathfrak s,\mathfrak f)}(\g_I)=0$ which shows that our result is independent of all these choices.

Now we take the limit where $B_a$ and $B_b$ go to zero that yields a simpler and more transparent expression which is manifestly independent of $S'_a$, $\F'_a$ or $B_a$. Let us focus on the contribution due to $\hat\p [-(B_a,\tilde \F_a)+ (B_a,\tilde\F_{\Pi_a}) ]$, namely
\be\label{rt}
j_{(\mathfrak S,\mathfrak F)}(\g_I)\supset \int_{B}\g_I\,\w\,(\tilde \F_{\Pi}-\tilde\F).
\ee
where we dropped the subscript $a$ to simplify the notation. Let us define $\H=\tilde \F_{\Pi}-\tilde\F$ which satisfies
\be\label{edp}
d\H=\d^3_{B}(\Pi),\qquad \H|_{S}=-\F,\qquad \H|_{S'}=0.
\ee
We have that $B=S\times I_L$ with $I_L$ an interval of length $L$ with coordinate $t\in[0,L]$ and a slicing of $B$ in $S_t$ with $S_0=S$ and $S_L=S'$. Since the integral above does not depend on $B$ it can not depend on $L$ so we may take the limit $L\rightarrow 0$ to make it manifestly independent of $B$. We define
\bea
\g_I&=&\g_I|_{S_t}+\tilde\g_I\\
\H&=&\H|_{S_t}+\tilde\H.
\eea
The equation for $\H$ translates into
\be\label{edp2}
d_S(\H|_{S_t})+d_{I_L}(\H|_{S_t})+d_S\tilde \H=\d^2_{S_0}(\Pi)\,\w\,\d(t)dt,\qquad \H|_{S_0}=-\F,\qquad \H|_{S_L}=0
\ee
where we used the fact that $d=d_{I_L}+d_{S}$ with $d_{I_L}$ and $d_S$ the exterior derivatives on $I_L$ and $S_t$ respectively. From the boundary conditions for $\H$ we find that
\be
\lim_{L\rightarrow 0}d_S(\H|_{S_t})=0,\qquad\lim_{L\rightarrow 0}d_{I_L}(\H|_{S_t})=\F\,\w\,\d(t)dt.
\ee
Thus, the differential equation for $L\rightarrow 0$ is
\be\label{edp3}
d_S\tilde \H=(\d^2_{S_0}(\Pi)-\F)\,\w\,\d(t)dt
\ee
so we necessarily have that
\be
\lim_{L\rightarrow 0}\tilde \H=-\frac{1}{2\pi l_s}A_\Pi\,\w\,\d(t)dt
\ee
with $A_\Pi$ a 1-form in $S$ that satisfies
\be\label{edp4}
\frac{1}{2\pi l_s}d_S A_\Pi=\F-\d^2_{S}(\Pi).
\ee
Going back to the integral (\ref{rt}) we find
\be
\int_{B}\g_I\,\w\,(\tilde \F_{\Pi}-\tilde\F)=\int_{S\times I_L}(\g_I|_{S_t}\,\w\,\tilde \H+\tilde\g_I\,\w\,\H|_{S_t})
\ee
where only the first term contributes in the limit $L\rightarrow 0$ since it contains a delta function unlike the second one. Therefore,%
\be
\int_{B}\g_I\,\w\,(\tilde \F_{\Pi}-\tilde\F)=-\frac{1}{2\pi l_s}\int_{S\times I_L}\g_I|_{S}\,\w\,\tilde A_\Pi\,\w\,\d(t)dt=\frac{1}{2\pi l_s}\int_{S}\g_I\,\w\,\tilde A_\Pi.
\ee
Notice there is minus sign due to the orientation of $S$ in $\p B=S'-S$.

Using this result we may write
\be\label{fin}
j_{(\mathfrak S,\mathfrak F)}(\g_I)=\int_{\Sig}\g_I+\frac{1}{2\pi l_s}\left(\int_{S_a}\g_I\,\w\,A_{\Pi_a}-\int_{S_b}\g_I\,\w\,A_{\Pi_b}\right).
\ee
The only thing left is to show that (\ref{fin}) is independent of the choice of $\Pi_a$ and $\Pi_b$. Consider $\Pi'_a=\Pi_a+\p\sigma_a$ and $\Pi'_b=\Pi_b+\p\sigma_b$ which also changes $\Sigma$ into $\Sigma'=\Sigma+\sigma_a-\sigma_b$. One can readily show that the formula above is independent of $\sigma_a$ and $\sigma_b$. Finally, a change $\Sigma'=\Sigma+\pi$ with $\pi$ a closed 3-cycle in $\mathcal M_6$ does change the expression above but just by integer number which can be interpreted as a redefinition of the $U(1)$ sector.

As a further check of this result let us derive eq.(\ref{charge}) starting from $j_{(\mathfrak S,\mathfrak F)}(\g_I)$ when $[F_a]=[F_b]\in H^2(S)$. More explicitly, we show that
\be
\int_\G\g\,\w\,\tilde \F=\int_\Sig\g+\int_{B_a}\g\,\w\,H_a-\int_{B_b}\g\,\w\,H_b.
\ee
Since nothing depends on $B_a$ and $B_b$ we choose $B_a=-B_b=-\G$. Also, since  $[F_a]=[F_b]\in H^2(S)$ we have that we may take $\Sig\subset\G$ so
\be
\int_\Sig\g+\int_{B_a}\g\,\w\,H_a-\int_{B_b}\g\,\w\,H_b=\int_\G\g\,\w\,(\d^2_\G(\Sig)-H_a+H_b).
\ee
The quantity $\Q\equiv\d^2_\G(\Sig)+H_a-H_b$ satisfies the equation
\be
d\Q=\d^3_{S_b}(\Pi_b)-\d_{S_a}^3(\Pi_a)+d\d^2_\G(\Sig)=0
\ee
where we used $d\d^2_\G(\Sig)=\d^3_{\G}(\p\Sig)=\d^3_{S_a}(\Pi_a)-\d_{S_b}^3(\Pi_b)$. The boundary conditions are $\Q|_{S_a}=\F_a$ and $\Q|_{S_b}=\F_b$ so $\Q=\tilde \F$.


\begin{thebibliography}{10}


\bibitem{thebook}
 L.~E.~Ib\'a\~nez and A.~M.~Uranga, 
  {\it String Theory and Particle Physics. An Introduction to String Phenomenology},
  Cambridge University Press (2012).

\bibitem{cchv09}
  S.~Cecotti, M.~C.~N.~Cheng, J.~J.~Heckman and C.~Vafa,
  {\em ``Yukawa Couplings in F-theory and Non-Commutative Geometry,''}
  [arXiv:0910.0477 [hep-th]].

\bibitem{fimr12}
  A.~Font, L.~E.~Ib\'a\~nez, F.~Marchesano and D.~Regalado,
  {\em ``Non-perturbative effects and Yukawa hierarchies in F-theory SU(5) Unification,''}
  JHEP {\bf 1303}, 140 (2013)
  [arXiv:1211.6529 [hep-th]].
  
\bibitem{fmrz13}
  A.~Font, F.~Marchesano, D.~Regalado and G.~Zoccarato,
  {\em ``Up-type quark masses in SU(5) F-theory models,''}
  JHEP {\bf 1311}, 125 (2013)
  [arXiv:1307.8089 [hep-th]].

\bibitem{Hattori:1993zu} 
  C.~Hattori, M.~Matsuda, T.~Matsuoka and D.~Mochinaga,
  {\em ``The String unification of gauge couplings and gauge kinetic mixings,''}
  Prog.\ Theor.\ Phys.\  {\bf 90}, 895 (1993)
  [hep-ph/9307305].

\bibitem{Dienes:1996zr} 
  K.~R.~Dienes, C.~F.~Kolda and J.~March-Russell,
  {\em``Kinetic mixing and the supersymmetric gauge hierarchy,''}
  Nucl.\ Phys.\ B {\bf 492}, 104 (1997)
  [hep-ph/9610479].

\bibitem{Lukas:1999nh} 
  A.~Lukas and K.~S.~Stelle,
  {\em ``Heterotic anomaly cancellation in five-dimensions,''}
  JHEP {\bf 0001}, 010 (2000)
  [hep-th/9911156].


 \bibitem{Lust:2003ky} 
  D.~L\"ust and S.~Stieberger,
  {\em ``Gauge threshold corrections in intersecting brane world models,''}
  Fortsch.\ Phys.\  {\bf 55}, 427 (2007)
  [hep-th/0302221].


\bibitem{Abel:2003ue} 
  S.~A.~Abel and B.~W.~Schofield,
  {\em ``Brane anti-brane kinetic mixing, millicharged particles and SUSY breaking,''}
  Nucl.\ Phys.\ B {\bf 685}, 150 (2004)
  [hep-th/0311051].

\bibitem{jl04}
  H.~Jockers and J.~Louis,
  {\em ``The Effective action of D7-branes in N = 1 Calabi-Yau orientifolds,''}
  Nucl.\ Phys.\ B {\bf 705}, 167 (2005)
  [hep-th/0409098].


\bibitem{Blumenhagen:2005ga} 
  R.~Blumenhagen, G.~Honecker and T.~Weigand,
  {\em ``Loop-corrected compactifications of the heterotic string with line bundles,''}
  JHEP {\bf 0506}, 020 (2005)
  [hep-th/0504232].

\bibitem{Abel:2006qt} 
  S.~A.~Abel, J.~Jaeckel, V.~V.~Khoze and A.~Ringwald,
  {\em ``Illuminating the Hidden Sector of String Theory by Shining Light through a Magnetic Field,''}
  Phys.\ Lett.\ B {\bf 666}, 66 (2008)
  [hep-ph/0608248].

\bibitem{agjkr08}
  S.~A.~Abel, M.~D.~Goodsell, J.~Jaeckel, V.~V.~Khoze and A.~Ringwald,
  {\em ``Kinetic Mixing of the Photon with Hidden $U(1)$s in String Phenomenology,''}
  JHEP {\bf 0807}, 124 (2008)
  [arXiv:0803.1449 [hep-ph]].
  
\bibitem{Grimm:2008dq} 
  T.~W.~Grimm, T.~-W.~Ha, A.~Klemm and D.~Klevers,
  {\em ``The D5-brane effective action and superpotential in N=1 compactifications,''}
  Nucl.\ Phys.\ B {\bf 816}, 139 (2009)
  [arXiv:0811.2996 [hep-th]].


\bibitem{Goodsell:2009xc} 
  M.~Goodsell, J.~Jaeckel, J.~Redondo and A.~Ringwald,
  {\em ``Naturally Light Hidden Photons in LARGE Volume String Compactifications,''}
  JHEP {\bf 0911}, 027 (2009)
  [arXiv:0909.0515 [hep-ph]].

\bibitem{Gmeiner:2009fb} 
  F.~Gmeiner and G.~Honecker,
  {\em ``Complete Gauge Threshold Corrections for Intersecting Fractional D6-Branes: The Z6 and Z6' Standard Models,''}
  Nucl.\ Phys.\ B {\bf 829}, 225 (2010)
  [arXiv:0910.0843 [hep-th]].


\bibitem{Goodsell:2010ie} 
  M.~Goodsell and A.~Ringwald,
  {\em ``Light Hidden-Sector $U(1)$s in String Compactifications,''}
  Fortsch.\ Phys.\  {\bf 58}, 716 (2010)
  [arXiv:1002.1840 [hep-th]].


\bibitem{Bullimore:2010aj} 
  M.~Bullimore, J.~P.~Conlon and L.~T.~Witkowski,
  {\em ``Kinetic mixing of $U(1)$s for local string models,''}
  JHEP {\bf 1011}, 142 (2010)
  [arXiv:1009.2380 [hep-th]].
  
  
\bibitem{Cicoli:2011yh} 
  M.~Cicoli, M.~Goodsell, J.~Jaeckel and A.~Ringwald,
  {\em ``Testing String Vacua in the Lab: From a Hidden CMB to Dark Forces in Flux Compactifications,''}
  JHEP {\bf 1107}, 114 (2011)
  [arXiv:1103.3705 [hep-th]].

\bibitem{Williams:2011qb} 
  M.~Williams, C.~P.~Burgess, A.~Maharana and F.~Quevedo,
  {\em ``New Constraints (and Motivations) for Abelian Gauge Bosons in the MeV-TeV Mass Range,''}
  JHEP {\bf 1108}, 106 (2011)
  [arXiv:1103.4556 [hep-ph]].

 \bibitem{gl11}
  T.~W.~Grimm and D.~V.~Lopes,
  {\em ``The N=1 effective actions of D-branes in Type IIA and IIB orientifolds,''}
  Nucl.\ Phys.\ B {\bf 855}, 639 (2012)
  [arXiv:1104.2328 [hep-th]].

\bibitem{kt11}
  M.~Kerstan and T.~Weigand,
  {\em ``The Effective action of D6-branes in N=1 type IIA orientifolds,''}
  JHEP {\bf 1106}, 105 (2011)
  [arXiv:1104.2329 [hep-th]].

\bibitem{cim11}
  P.~G.~C\'amara, L.~E.~Ib\'a\~nez and F.~Marchesano,
  {\em ``RR photons,''}
  JHEP {\bf 1109}, 110 (2011)
  [arXiv:1106.0060 [hep-th]].


\bibitem{Honecker:2011sm} 
  G.~Honecker,
  {\em ``Kaehler metrics and gauge kinetic functions for intersecting D6-branes on toroidal orbifolds - The complete perturbative story,''}
  Fortsch.\ Phys.\  {\bf 60}, 243 (2012)
  [arXiv:1109.3192 [hep-th]].

\bibitem{Goodsell:2011wn} 
  M.~Goodsell, S.~Ramos-Sanchez and A.~Ringwald,
  {\em ``Kinetic Mixing of $U(1)$s in Heterotic Orbifolds,''}
  JHEP {\bf 1201}, 021 (2012)
  [arXiv:1110.6901 [hep-th]].

\bibitem{Honecker:2012qr} 
  G.~Honecker, M.~Ripka and W.~Staessens,
  {\em ``The Importance of Being Rigid: D6-Brane Model Building on $T^6/Z_2 x Z_6'$ with Discrete Torsion,''}
  Nucl.\ Phys.\ B {\bf 868}, 156 (2013)
  [arXiv:1209.3010 [hep-th]].

\bibitem{Shiu:2013wxa} 
  G.~Shiu, P.~Soler and F.~Ye,
  {\em ``Millicharged Dark Matter in Quantum Gravity and String Theory,''}
  Phys.\ Rev.\ Lett.\  {\bf 110}, no. 24, 241304 (2013)
  [arXiv:1302.5471 [hep-th]].

\bibitem{Holdom:1985ag} 
  B.~Holdom,
  {\em ``Two $U(1)$'s and Epsilon Charge Shifts,''}
  Phys.\ Lett.\ B {\bf 166}, 196 (1986).

\bibitem{Feldman:2006wd}
  D.~Feldman, B.~Kors, P.~Nath,
  {\em ``Extra-weakly Interacting Dark Matter,''}
  Phys.\ Rev.\  {\bf D75 } (2007)  023503.
  [hep-ph/0610133];

\bibitem{Feldman:2009wv}
  D.~Feldman, Z.~Liu, P.~Nath, B.~D.~Nelson,
  {\em ``Explaining PAMELA and WMAP data through Coannihilations in Extended SUGRA with Collider Implications,''}
  Phys.\ Rev.\  {\bf D80 } (2009)  075001.
  [arXiv:0907.5392 [hep-ph]].

\bibitem{Ibarra:2008kn}
  A.~Ibarra, A.~Ringwald, C.~Weniger,
  {\em ``Hidden gauginos of an unbroken $U(1)$: Cosmological constraints and phenomenological prospects,''}
  JCAP {\bf 0901 } (2009)  003.
  [arXiv:0809.3196 [hep-ph]].


\bibitem{Arvanitaki:2009hb}
  A.~Arvanitaki, N.~Craig, S.~Dimopoulos, S.~Dubovsky, J.~March-Russell,
  {\em ``String Photini at the LHC,''}
  Phys.\ Rev.\  {\bf D81 } (2010)  075018.
  [arXiv:0909.5440 [hep-ph]].


\bibitem{Hitchin99} 
  N.~J.~Hitchin,
  {\em ``Lectures on special Lagrangian submanifolds,''}
  math/9907034.
 
\bibitem{gl04}
  T.~W.~Grimm, J.~Louis,
  {\em ``The Effective action of type IIA Calabi-Yau orientifolds,''}
  Nucl.\ Phys.\  {\bf B718 } (2005) 153-202.
  [hep-th/0412277].
 
 
\bibitem{reviews}
  R.~Blumenhagen, M.~Cveti\v c, P.~Langacker, G.~Shiu,
  {\em ``Toward realistic intersecting D-brane models,''}
  Ann.\ Rev.\ Nucl.\ Part.\ Sci.\  {\bf 55 } (2005)  71-139.
  [hep-th/0502005];

  R.~Blumenhagen, B.~K\"ors, D.~L\"ust, S.~Stieberger,
  {\em ``Four-dimensional String Compactifications with D-Branes, Orientifolds and Fluxes,''}
  Phys.\ Rept.\  {\bf 445 } (2007)  1-193.
  [hep-th/0610327];

  F.~Marchesano,
  {\em ``Progress in D-brane model building,''}
  Fortsch.\ Phys.\  {\bf 55 } (2007)  491-518.
  [hep-th/0702094 [HEP-TH]];

 \bibitem{Witten79} 
  E.~Witten,
  {\em ``Dyons of charge $e\th/2\pi$,''}
  Phys. Lett. B 89 (1979).

 
  \bibitem{bjk09}
  F.~Br\"ummer, J.~Jaeckel and V.~V.~Khoze,
  {\em ``Magnetic mixing,''}
  JHEP {\bf 0906} 037  (2009)
  [arXiv:0905.0633 [hep-ph]].


\bibitem{dps14}
  M.~R.~Douglas, D.~S.~Park and C.~Schnell,
  {\em ``The Cremmer-Scherk Mechanism in F-theory Compactifications on K3 Manifolds,''}
  arXiv:1403.1595 [hep-th].

\bibitem{Beasley:2008kw}
  C.~Beasley, J.~J.~Heckman, C.~Vafa,
  {\em ``GUTs and Exceptional Branes in F-theory - II: Experimental Predictions,''}
  JHEP {\bf 0901 } (2009)  059.
  [arXiv:0806.0102 [hep-th]].

   \bibitem{dw08}
  R.~Donagi, M.~Winhjolt,
  {\em ``Breaking GUT groups in F-theory,''}
  Adv. Theor. Math. Phys. 15 (2011) 1523-1603
  [arXiv:0808.2223 [hep-th]].

\bibitem{deWit:1984px} 
  B.~de Wit, P.~G.~Lauwers and A.~Van Proeyen,
  {\em ``Lagrangians of N=2 Supergravity - Matter Systems,''}
  Nucl.\ Phys.\ B {\bf 255}, 569 (1985).

\bibitem{Andrianopoli:1996cm} 
  L.~Andrianopoli, M.~Bertolini, A.~Ceresole, R.~D'Auria, S.~Ferrara, P.~Fre and T.~Magri,
  {\em ``N=2 supergravity and N=2 superYang-Mills theory on general scalar manifolds: Symplectic covariance, gaugings and the momentum map,''}
  J.\ Geom.\ Phys.\  {\bf 23}, 111 (1997)
  [hep-th/9605032].
  
  \bibitem{Grimm:2004uq} 
  T.~W.~Grimm and J.~Louis,
  {\em ``The Effective action of N = 1 Calabi-Yau orientifolds,''}
  Nucl.\ Phys.\ B {\bf 699}, 387 (2004)
  [hep-th/0403067].

\bibitem{Buican:2006sn}
  M.~Buican, D.~Malyshev, D.~R.~Morrison, H.~Verlinde, M.~Wijnholt,
  {\em ``D-branes at Singularities, Compactification, and Hypercharge,''}
  JHEP {\bf 0701 } (2007) 107.
  [hep-th/0610007].

\bibitem{em07}
  J.~Evslin, L.~Martucci,
  {\em ``D-brane networks in flux vacua, generalized cycles and calibrations,''}
  JHEP {\bf 0707} 040  (2007) 
  [arXiv:0703.129 [hep-th]].

 \bibitem{Redondo:2008zf} 
  J.~Redondo,
  {\em ``The Low energy frontier: Probes with photons,''}
  arXiv:0805.3112 [hep-ph].
  
  \bibitem{Donagi:2008ca}
  R.~Donagi and M.~Wijnholt,
  {\em ``Model Building with F-Theory,''}
  Adv.\ Theor.\ Math.\ Phys.\  {\bf 15}, 1237 (2011)
  [arXiv:0802.2969 [hep-th]].

\bibitem{Blumenhagen:2008aw}
  R.~Blumenhagen,
  {\em ``Gauge Coupling Unification in F-Theory Grand Unified Theories,''}
  Phys.\ Rev.\ Lett.\  {\bf 102 } (2009)  071601.
  [arXiv:0812.0248 [hep-th]].

\bibitem{Conlon:2009qa}
  J.~P.~Conlon and E.~Palti,
  {\em ``On Gauge Threshold Corrections for Local IIB/F-theory GUTs,''}
  Phys.\ Rev.\  D {\bf 80} (2009) 106004
  [arXiv:0907.1362 [hep-th]].

  
\bibitem{ruback86} 
  P.~J.~Ruback,
  {\em ``The motion of Kaluza-Klein monopoles,''}
  Commun. Math. Phys. 107 (1979) 93-102.


  
 
  
\end{thebibliography}
\end{document}